\documentclass[preprint,showpacs,preprintnumbers,amsmath,amssymb,prd]{revtex4} 
\usepackage{amsmath}
\usepackage{setspace}
\usepackage[margin=10pt, font={footnotesize}, labelfont=bf, labelsep=endash, justification=raggedright]{caption}
\usepackage{array}
\usepackage{comment}
\usepackage{graphicx,subfigure}
\usepackage{dcolumn}
\usepackage{bm}
\usepackage{slashed}
\usepackage[hidelinks]{hyperref}
\usepackage{float}
\newcommand{\be}{\begin{eqnarray}}
\newcommand{\ee}{\end{eqnarray}}
\newcommand{\bea}{\begin{eqnarray}}
\newcommand{\eea}{\end{eqnarray}}

\newif\ifdraft
\draftfalse

\begin{document}
\title{Raising the Higgs Mass in Supersymmetry with $t-t'$ Mixing}
\author{Cyrus Faroughy}
\email{cfarough@pha.jhu.edu}
\affiliation{Department of Physics and Astronomy, Johns Hopkins University, Baltimore, MD 21218}
\author{Kevin Grizzard}
\email{kgrizz@pha.jhu.edu}
\affiliation{Department of Physics and Astronomy, Johns Hopkins University, Baltimore, MD 21218}


\begin{abstract}
In this article we propose a new strategy to address the Little Hierarchy problem. We show that the addition of a fourth generation with vector-like quarks to the minimal supersymmetric standard model (MSSM) can raise the predicted value of the physical Higgs mass by mixing with the top sector. The mixing requires a larger top quark Yukawa coupling (by up to $\sim 6 \%$) to produce the same top mass. Since loop corrections to $m_h$ go as $y_{top}^4$, this will in turn increase the predicted value of the physical Higgs mass, a point not previously emphasized in the literature. In the presence of mixing, for $A$-terms and soft masses around 900 GeV, a Higgs mass of 125 GeV can be generated while  retaining perturbativity of the gauge couplings, evading constraints from electroweak precision measurements (EWPM) and recent LHC searches, and pushing the Landau pole for the top Yukawa above the GUT scale.  Soft masses can be as low as 800 GeV in parts of parameter space with a Landau pole at $\sim 10^{10}$ GeV. However, the Landau pole can still be pushed above the GUT scale if one sacrifices perturbative unification by adding fields in a $\mathbf{5}$+$\mathbf{\bar{5}}$ representation. With a ratio of weak-scale vector masses $\neq 1$, soft masses may be slightly below $800$ GeV. The model predicts new quarks and squarks with masses $\gtrapprox 750$ GeV. We briefly discuss potential paths for discovery or exclusion at the LHC.
\end{abstract}

\maketitle
\tableofcontents
\section{Introduction}
Dynamically broken supersymmetry offers an elegant way of cutting off leading divergences of quantum corrections to the Higgs mass parameter in the standard model.  Unless parameters in the model are finely tuned, one expects that the mass of supersymmetric particles are of the same order as the $Z$ and $W$ masses. In particular, quantum corrections to the Higgs mass parameter are dominated by the contributions from the top quark, because of the large Yukawa coupling. To preserve naturalness, this leads to the expectation that the top squark should be relatively light.

However, results from the Large Hadron Collider (LHC) indicate that the Higgs mass is $\sim$ 125 GeV \cite{ATLAShiggs:2012, CMShiggs:2012}. In the MSSM, a mass so much higher than the tree-level upper bound of $m_Z$ can be accommodated only with extremely heavy top squarks, or moderately heavy top squarks and large top squark mixing. The quadratic divergence contributed by such a heavy top squark then needs to be cancelled at the level of $\sim 10^{-4}$,  leading to a significantly fine-tuned theory. This tuning is significantly worse than the tuning implied by direct constraints on superpartners at the LHC. In fact, in the case of only moderate mixing, the top squark mass implied by this Higgs mass is higher than the direct collider limit $\sim 3$ TeV that can ever be set by the LHC.

Unlike many other experimental constraints on the MSSM, this ``Little Hierarchy" problem   \cite{Barbieri:2000gf, Giudice:2006sn} is directly associated with the low energy spectrum of the theory. Consequently, it cannot be solved through ultraviolet mechanisms that are often invoked to address indirect constraints (such as flavor or CP violation, see \cite{Luty:2005sn} for an overview)  or alteration of the collider signatures of supersymmetry to avoid direct constraints on the theory   \cite{Carpenter:2007zz, Carpenter:2008sy, Graham:2012th, Fan:2011yu, Kribs:2012gx, Baryakhtar:2012rz}. Several attempts have been made to modify the MSSM spectrum through the addition of matter fields to raise the Higgs mass \cite{Choi:2005hd, Kitano:2005wc, Chacko:2005ra, Ellis:1988er, Espinosa:1998re, Batra:2003nj, Maloney:2004rc, Casas:2003jx, Brignole:2003cm, Harnik:2003rs, Chang:2004db, Delgado:2005fq, Birkedal:2004zx, Babu:2004xg,  Choi:2006xb, Dutta:2007az, Abe:2007je, Dermisek:2006ey, Abe:2007kf, Dutta:2007xr, Kikuchi:2008ws, Kim:2006mb, Dine:2007xi, Dermisek:2005ar, Bellazzini:2009ix,  Gogoladze:2009bd,  Graham:2009gy, Martin:2009bg,  Burdman:2006tz, Chacko:2005pe, Chang:2006ra, Falkowski:2006qq, ArkaniHamed:2002qy, ArkaniHamed:2001nc}. These mechanisms were originally proposed to accommodate the Higgs mass bound $\gtrapprox 114$ GeV imposed by LEP, and though more recent work has demonstrated the ability for such a mechanism to yield a Higgs mass $\sim 125$ GeV (e.g., \cite{Martin:2012dg}), in general the higher mass needs significantly larger couplings than considered in the earlier models, leading to the rapid appearance of Landau poles marginally above the weak scale. While such a possibility cannot be logically excluded, it destroys the success of perturbative grand unification in supersymmetric models, an aesthetic success of the MSSM.

In this paper, we propose a new strategy to address the Little Hierarchy problem. The largest loop contribution to the effective potential of the Higgs comes from the top supermultiplet and the magnitude of this contribution is governed by the top Yukawa. The Yukawa coupling used in current estimates of the top quark contribution to the Higgs mass is directly extracted from measurements of the top mass. However, the naive relation between the physical mass of the top quark and the Yukawa coupling, extracted from the tree level Lagrangian, is modified when the top supermultiplet is mixed with other heavier states. When diagonalizing the mass matrix, the new mixing terms will contribute negatively to the naive estimate $y_t v\sin\beta$, thus requiring a larger Yukawa coupling to obtain the measured value of the top, $m_t\sim173$ GeV. Since the Higgs effective potential depends upon the fourth power of this coupling, even a moderate increase can lead to a significant enhancement of the Higgs mass.

We demonstrate this mechanism through a simple extension of the models \cite{Graham:2009gy, Martin:2012dg, Martin:2009bg} where a vector-like fourth generation with Yukawa couplings to the Higgs was introduced. In these models, the additional contributions from the vector-like generation was sufficient to push the Higgs mass above the LEP bound of $\sim 114$ GeV. This goal could be accommodated with perturbative gauge coupling unification with relative ease using only the Yukawa couplings of the fourth generation with itself. Consequently, mixing between the fourth generation and the standard model was not explored. But the mixing between the top quark and the fourth generation is experimentally fairly unconstrained. Indeed, recently there has been more interest shown in exploring this possibility, with \cite{VLQ:2013} in particular seeking to constrain the possible dominant mixing angle for any (single) vector-like heavy multiplet.  However, it has not been noted that such a mixing can contribute significantly to the mechanism for raising $m_h$ so far above $m_Z$. When this mixing is $\mathcal{O}\left(1\right)$, we show that the Yukawa couplings necessary to obtain the physical top quark mass are large enough to substantially increase the Higgs mass.
\\
\indent This paper is structured as follows. We describe the model in section \ref{Sec:Model}. In section \ref{sec:EffectsfromMixing}, we discuss the effects of large mixing on the top Yukawa. We compute the weak-scale mixing Yukawa couplings necessary to achieve a Higgs mass of $\sim$ 125 GeV and the induced top Yukawa Landau pole. In section \ref{Sec:Constraints} we study the experimental constraints and briefly discuss the LHC phenomenology. Finally, we conclude in section \ref{sub:conclusion}.

\section{The Model}
\label{Sec:Model}
In this model, we extend the MSSM by adding a full vector-like fourth generation (i.e., a chiral fourth generation plus its mirror) with Yukawa couplings to the Higgs. Furthermore, the couplings mixing the fourth generation and the top sector are allowed to take on values close to unity; they have a quasi-fixed point which limits their TeV values to be not much larger than 1 \cite{Martin:2009bg}. However, we ignore mixing with the first and second generations since these are constrained by experiment to be small. We consider the simplest model which preserves gauge coupling unification. Therefore, the new vector-like generation contains quark and lepton supermultiplets $Q_4$, $U^{c}_4$ and $E^{c}_4$, living in the $\mathbf{10}$ representation of SU(5), plus the corresponding mirror generation $\bar{Q}^{c}_4$, $\bar{U}_4$, and $\bar{E}_4$ living in the $\mathbf{\bar{10}}$ representation. The $SU(3)_c \times SU(2)_L \times U(1)_Y$ quantum numbers of the additional coloured superfields and the top sector, plus explanation of our conventions and notation are shown in Table \ref{Tab:Table1}.
\begin{table}
\renewcommand*\arraystretch{1.0}
  \begin{tabular}{|c|c|c||c|c|c|c|c|c|}
    \hline
    \textbf{Supermultiplet} & \textbf{Scalars } & \textbf{Fermions} & $\mathbf{SU(3)_{C}}$ & $\mathbf{SU(2)_{L}}$ & $\mathbf{U(1)_{Y}}$ & $\mathbf{T^{3}}$ & $\mathbf{Q}$ \\\hline\hline
    $Q_3$ & $(\tilde{u_3}, \tilde{d_3})$ & $(u_3, d_3)$ & $\mathbf{3}$ & $\mathbf{2}$ & 1/6 & (1/2,-1/2) & (2/3,-1/3) \\[1ex]\hline
    $U^{c}_3$ & $\tilde{u}^c_3$ & $u^c_3$ & $\mathbf{\bar{3}}$ & $\mathbf{1}$ & -2/3 & 0 & -2/3 \\[1ex]\hline
    $D^{c}_3$ & $\tilde{d}^{c}_3$ & $d^c_3$ & $\mathbf{\bar{3}}$ & $\mathbf{1}$ & 1/3 & 0 & 1/3 \\[1ex] \hline\hline
    $Q_4$ & $(\tilde{u}_{4}, \tilde{d}_{4})$ & $(u_{4}, d_{4})$ & $\mathbf{3}$ & $\mathbf{2}$ & 1/6 & (1/2,-1/2) & (2/3,-1/3) \\[1ex]\hline
    $U^{c}_4$ & $\tilde{u}^c_{4}$ & $u^c_{4}$ & $\mathbf{\bar{3}}$ & $\mathbf{1}$ & -2/3 & 0 & -2/3\\[1ex]\hline
    $\bar{Q}^{c}_4$ & $(\tilde{\bar{d}}^{c}_{4}, \tilde{\bar{u}}^{c}_{4})$ & $(\bar{d}^c_{4}, \bar{u}^c_{4})$ & $\mathbf{\bar{3}}$ & $\mathbf{2}$ & -1/6 & (1/2,-1/2) & (1/3,-2/3) \\[1ex]\hline
    $\bar{U}_4$ & $\tilde{\bar{u}}_{4}$ & $\bar{u}_{4}$ & $\mathbf{3}$ & $\mathbf{1}$ & 2/3 & 0 & 2/3 \\[1ex]
    \hline
\end{tabular}
  \caption[]{The third and fourth generation coloured fields and their quantum numbers in the gauge eigenstate basis are listed in the table above. We follow the standard convention that all chiral supermultiplets are defined in terms of 2-component left-handed Weyl spinors, so that charge conjugates of right-handed fields are used. The barred fields denote gauge-eigenstate fields belonging to the $\mathbf{\bar{10}}$ representation of $SU(5)$. 4-component Dirac fermions can be constructed as $q_D=(q_i, q_i^{c\dagger})^T$. The mass basis fermions are the top $t$, bottom $b$, and the new quarks $t'_{1,2}$ and $b'$ of charge +2/3 and -1/3, respectively. Their superparters are the top squarks $\tilde{t}_{1,2}$, bottom squarks $\tilde{b}_{1,2}$, and the corresponding non-MSSM squarks $\tilde{t}'_{1,2,3,4}$, and $\tilde{b}'_{1,2}$.}
  \label{Tab:Table1}
\end{table}

The relevant mass-eigenstate Dirac fermions are the top $t$, bottom $b$, and the new quarks $t'_{1,2}$ and $b'$ of charge +2/3 and -1/3, respectively. In the scalar sector the relevant particles are the top squarks $\tilde{t}_{1,2}$, bottom squarks $\tilde{b}_{1,2}$, and the corresponding non-MSSM squarks $\tilde{t}'_{1,2,3,4}$, and $\tilde{b}'_{1,2}$. The terms in the superpotential that affect the Higgs mass are:
\begin{equation}\label{Eq:Superpotential}
W \subset y_{ij} Q_i H_u U^{c}_j + \mu_Q \bar{Q}^{c}_4 Q_4 + \mu_U \bar{U}_4 U^{c}_4 +\mu H_u H_d
\end{equation}
where $i$ and $j$ are generation indices than run from 3 to 4, and $\mu$ is the usual coefficient of the Higgs bilinear term. Terms such as $\mu_{34} Q_3 \bar{Q}^{c}_4$ are rotated away without loss of generality. Yukawa couplings of the form $\bar{y}_{44} H_d \bar{Q}^c_4 \bar{U}_4$ and Yukawa couplings between the Higgs and the leptons are ignored since their effect in raising the Higgs mass is subdominant in the large $\tan\beta$ limit. In the soft Lagrangian, we assume the same squared mass $\Delta m^2$ for all the squarks, $B\mu$ terms corresponding to each vector-like mass (ignoring mixed $B\mu$ terms with the third generation), and $A$-terms of the form $ y_{ij} A$ associated with each Yukawa coupling. Throughout the paper, we set $\tan\beta=30$.  We refer to the appendices for details about the particle spectrum and the interaction Lagrangian.

\section{The Effects from Mixing}\label{sec:EffectsfromMixing}

\subsection{Mixing and the Top Yukawa Coupling}\label{sec:mixing}
As stated in the introduction, the qualitative difference between this note and earlier work \cite{Graham:2009gy, Martin:2012dg, Martin:2009bg} is the emphasis on the mixing terms proportional to $y_{34}$ and $y_{43}$. In general, we assume a parameter space where $y_{34}$, $y_{43}$ and $y_{44}$ are allowed to vary from 0 to values $\gtrsim1$, while the top Yukawa is constrained to give the right top mass. We consider the four following benchmark scenarios for the Yukawas: (1) $y_{34} = -y_{43} \gg y_{44}$, (2) $y_{43} \gg y_{34}, y_{44}$, (3) $y_{34} \gg y_{43}, y_{44}$, and (4) $y_{44} \gg y_{34}, y_{43}$. 
Case 1 focuses on effects where both mixing Yukawas are significant, whereas cases 2 and 3 focus on mixing from only one term. Case 4 corresponds to earlier work \cite{Graham:2009gy, Martin:2012dg, Martin:2009bg} where the mixing terms $y_{34}$ and $y_{43}$ were ignored, and serves as a useful comparison. As will be shown in section \ref{Sec:HiggsMass}, the parameter space where this model makes sizeable contributions to the Higgs mass is a region where the fourth generation is accessible at the LHC.\\
\indent When mixing terms are present, and if $y_{44}=0$, the top Yukawa coupling $y_{33}$ necessary to obtain the measured top mass $m_t=172.9$ GeV is given by:

\begin{equation}
y_{33} =\frac{m_t}{v\, \text{sin}\beta} \left(1+\frac{\left( y_{43} \, v \,\text{sin}\beta\right)^2}{\mu_Q^2 - m_t^2} \right)^{1/2}\left(1+\frac{\left( y_{34} \, v \,\text{sin}\beta\right)^2}{\mu_U^2 - m_t^2} \right)^{1/2}.
\label{Eqn:TopYukawa}
\end{equation}

This formula is exact when $y_{44}=0$ and is obtained after bi-diagonalizing the up-type fermion mass matrix $m_f^u$ (shown explicitly in appendix \ref{app:Spectrum}), identifying its smallest singular value with the top mass, and solving for $y_{33}$. If $y_{44}\ne0$, the above formula still holds to a very good approximation since the coupling $y_{44}$ first makes an appearance at fourth order in the expansion parameter ($v/\mu_{Q,U}$), and therefore has a negligible effect in raising the value of $y_{33}$. \\
\indent For simplicity, we take $\mu_Q=\mu_U\equiv\mu_4$. In this case, we can define $\Delta=v/\mu_4$ to quantify the hierarchy between the new vector-like mass scale and the electroweak scale, such that $\Delta=0$ in the limit $\mu_4\rightarrow \infty$. At large $\tan\beta$, and taking $m_t/v=1$, equation \ref{Eqn:TopYukawa} can be approximated as

\begin{equation}
y_{33} \approx 1+\frac{1}{2}\left(\frac{\Delta^2}{1 - \Delta^2}\right)\left( y_{43}^2+y_{34}^2\right) + \mathcal{O}(\Delta^4).
\label{Eqn:TopYukawa2}
\end{equation}
\indent Evidently, $\Delta>0$ leads to an increase in the top Yukawa. As a result, the soft masses $\Delta m$ needed to get a 125 GeV Higgs decrease. Taking the value of the mass of the new quarks to be near their experimental limit of $~700-800$ GeV (see section \ref{subsection:Massbounds}) leads to the constraint $\Delta\lesssim1/4$. Then, in the case where the mixing Yukawas are near unity, the effects of mixing between the top sector and the fourth generation can lead to an increase of $y_{33}$ by about $6 \%$. This can significantly increase the Higgs mass squared since the radiative corrections go as $y_{33}^4$. Mixing effects on the Higgs mass are studied in detail in section \ref{Sec:HiggsMass}. Lastly, we note that an increase in the top Yukawa also leads to an increase in the Higgs quartic; however, this increase is  subdominant compared to the Higgs mass.

\subsection{Weak-Scale Yukawa Couplings}\label{Sec:HiggsMass}
In this section we compute the weak-scale Yukawa couplings necessary to obtain the required Higgs mass using the one-loop effective potential in the decoupling limit (where $m_A, m_{H^+}, m_{H^-}, m_{H^0}>>m_{h}$). Contributions to the Higgs effective potential have the following form:

\begin{equation}
\Delta V= \frac{3}{32\pi^2}[\sum_{\{\tilde{m}_{a}\}}\tilde{m}^2_a\left(\ln{\frac{\tilde{m}^2_a}{Q^2}}-\frac{3}{2}\right) - 2\sum_{\{m_{a}\}}m^2_a\left(\ln{\frac{m^2_a}{Q^2}}-\frac{3}{2}\right)]
\end{equation}

\noindent where $Q$ is the renormalization scale and $m_a$ ($\tilde{m}_a$) are the quark (squark) masses. The summation runs over the masses of the heavy up-type quarks ($a=t, t'_1, t'_2$) and their superpartners ($a=\tilde{t}_{1,2}, \tilde{t}'_{1,2,3,4}$). The resulting physical Higgs mass is then

\begin{equation}
m_h= \sqrt{m^2_Z\cos^2{2\beta}+\frac{1}{2}\left(\frac{\partial^2(\Delta V)}{\partial v_u^2}-\frac{1}{v_u}\frac{\partial(\Delta V)}{\partial v_u}\right)}.
\end{equation}

\begin{figure}
\includegraphics[width = 5.0 in]{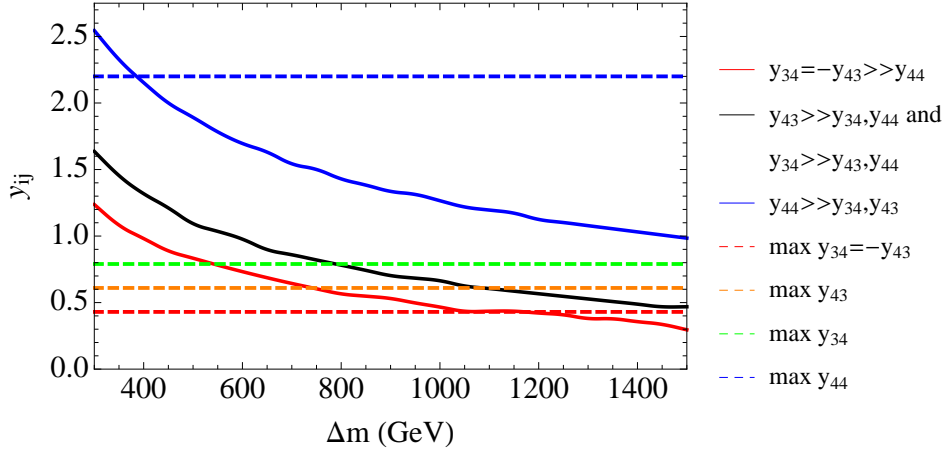}
 \caption{ We plot the values of the Yukawa couplings at the weak scale necessary to obtain $m_h= 125.5\pm0.5$ GeV, as a function of $\Delta m$. We take $A=\Delta m$, $\mu_4 = 900$ GeV. When either $y_{34}$ or $y_{43}$ dominates, the same value of the dominant Yukawa is required to get $m_h=125.5$ GeV so both scenarios are represented by one black line. The dotted lines show the maximum values allowed by EWPM for each mixing scenario (see section \ref{Sec:Constraints}). Since $y_{34}$ and $y_{43}$ contribute to the oblique parameters differently they have different constraints on their maximum values, represented by the green and orange dotted lines, respectively. Above the dotted line requires Yukawas larger than allowed by EWPM and is thus ruled out.  \label{Fig:yvalues} }
\end{figure}

For numerical efficiency, the algorithm used to solve for the necessary parameters obtains a Higgs mass in the range $125.5 \pm 0.5$ GeV.  For this set of computations we take the soft terms to be of the form $\Delta m = A$, as might be expected in gravity mediation (or high scale gauge mediation \cite{Graham:2009gy}), and choose $\mu_4 = 900$ GeV. The Yukawa values at the weak scale as functions of the soft masses are plotted in Figure \ref{Fig:yvalues}, along with their constraints from electroweak precision measurements.
As one would intuitively expect, the mixing Yukawas necessary to achieve a given Higgs mass are smaller when $ |y_{34}| \sim |y_{43}|$ than when one of these couplings dominates the other. However, the lowest possible value of $\Delta m$ consistent with EWPM is $\Delta m\sim 800$ GeV and occurs for the case where $y_{34}\sim0.8$ and $y_{43}=y_{44}=0$.

\subsection{Top Yukawa Landau Pole}\label{subsection:LandauPoles}
The mixing terms $y_{34}$ and $y_{43}$ significantly affect the Higgs mass only when they are $\mathcal{O}\left(1\right)$. These $\mathcal{O}\left(1\right)$ Yukawas affect the renormalization group evolution of the top Yukawa $y_{33}$ and can cause it to hit a Landau pole. In this section, we estimate the scale at which this Landau pole is attained for various choices of the Yukawas and soft terms necessary to obtain a Higgs mass $\sim$ 125 GeV. The top Yukawa two-loop beta function presented in appendix \ref{app:BetaFunctionsYuk} is used to calculate the scale $\Lambda$ where the coupling $y_{33}$ hits a Landau pole. Below, we plot $\Lambda$ as a function of the soft mass $\Delta m$ and consider the effects from:
\begin{enumerate}
\item Different mixing scenarios.
\item $A$-terms.
\item The vector-like mass $\mu_4$.
\item The number of extra multiplets in the ${\bf 5 } + {\bf \bar{5}}$ of SU(5).
\end{enumerate}
\begin{figure}
\includegraphics[width = 5.0 in]{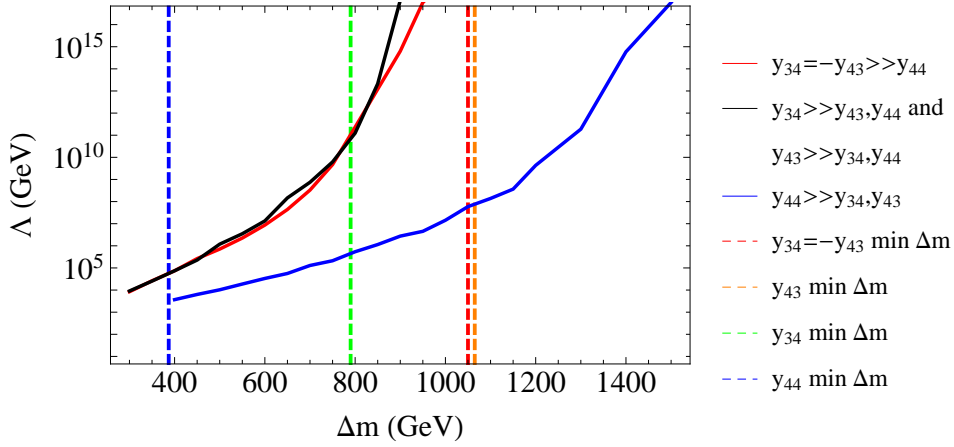}
\caption{We plot the scale $\Lambda$ where the $y_{33}$ required to get $m_h=125.5$ GeV hits a Landau pole, as a function of the soft mass $\Delta m$. We set $A=\Delta m$, $\mu_4 = 900$ GeV, and $n_5 = 0$. Soft masses to the left of the dotted lines can only yield $m_h=125.5$ GeV with Yukawa couplings larger than allowed by EWPM and are thus ruled out (see section \ref{Sec:Constraints}).  Physically uninteresting values of $\Lambda < 1$ TeV are not plotted.  The presence of mixing decreases significantly the value of the soft masses needed. As can be seen from the plot, the scale of the Landau pole in the cases with sizeable mixing are all comparable. The case where either $y_{34}$ or $y_{43}$ dominate (shown in black) yield identical values since each contributes to the top Yukawa beta function in the same way. However, their differing effects on the oblique parameters lead to different minimum values for the soft masses.
\label{Fig:yscales} }
\end{figure}
\begin{figure}
\includegraphics[width = 5.0 in]{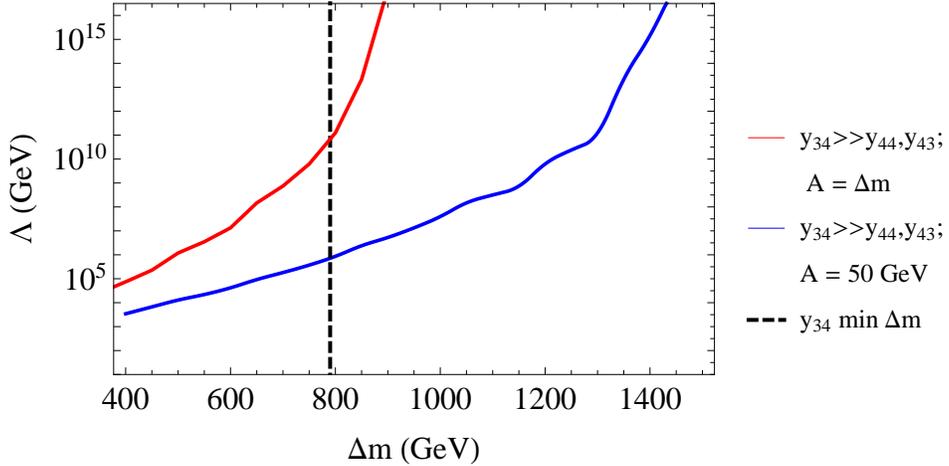}
\caption{We plot the scale $\Lambda$ where the $y_{33}$ required to get $m_h=125.5$ GeV hits a Landau pole, as a function of the soft mass $\Delta m$. We set $y_{34} \gg y_{44}, y_{43}$, $\mu_4 = 900$ GeV, and $n_5 = 0$. Soft masses to the left of the dotted lines can only yield $m_h=125.5$ GeV with Yukawa couplings larger than allowed by EWPM and are thus ruled out (see section \ref{Sec:Constraints}). There is only one line here since these limits are independent of the $A$-terms). For a given soft mass the implied Landau pole gets significantly pushed up by the presence of $A$-terms. \label{Fig:Aterms34}}
\end{figure}
\begin{figure}
\includegraphics[width = 5.0 in]{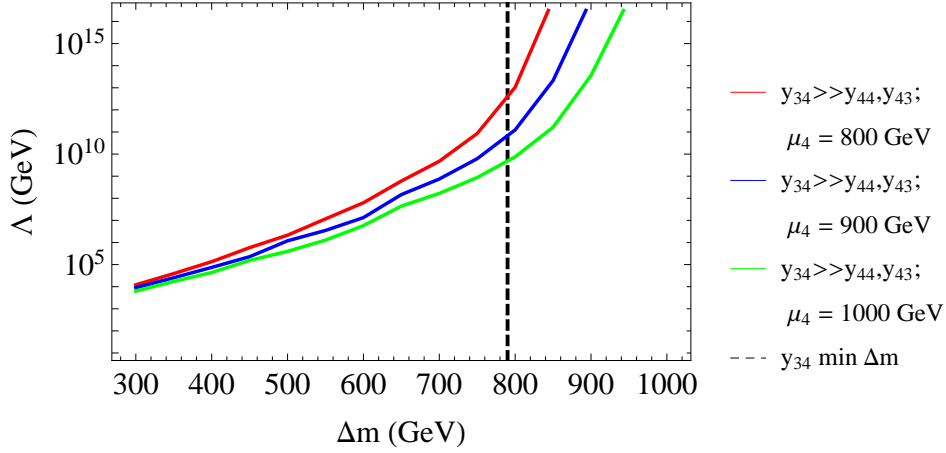}
\caption{We plot the scale $\Lambda$ where the $y_{33}$ required to get $m_h=125.5$ GeV hits a Landau pole, as a function of the soft mass $\Delta m$. We set $y_{34} \gg y_{44}, y_{43}$, $A=\Delta m$ , and $n_5 = 0$. Soft masses to the left of the dotted lines can only yield $m_h=125.5$ GeV with Yukawa couplings larger than allowed by EWPM and are thus ruled out (see section \ref{Sec:Constraints}). For a given soft mass, the implied Landau pole increases as the vector mass decreases. \label{Fig:vecplots} }
\end{figure}
\begin{figure}
\includegraphics[width = 5.0 in]{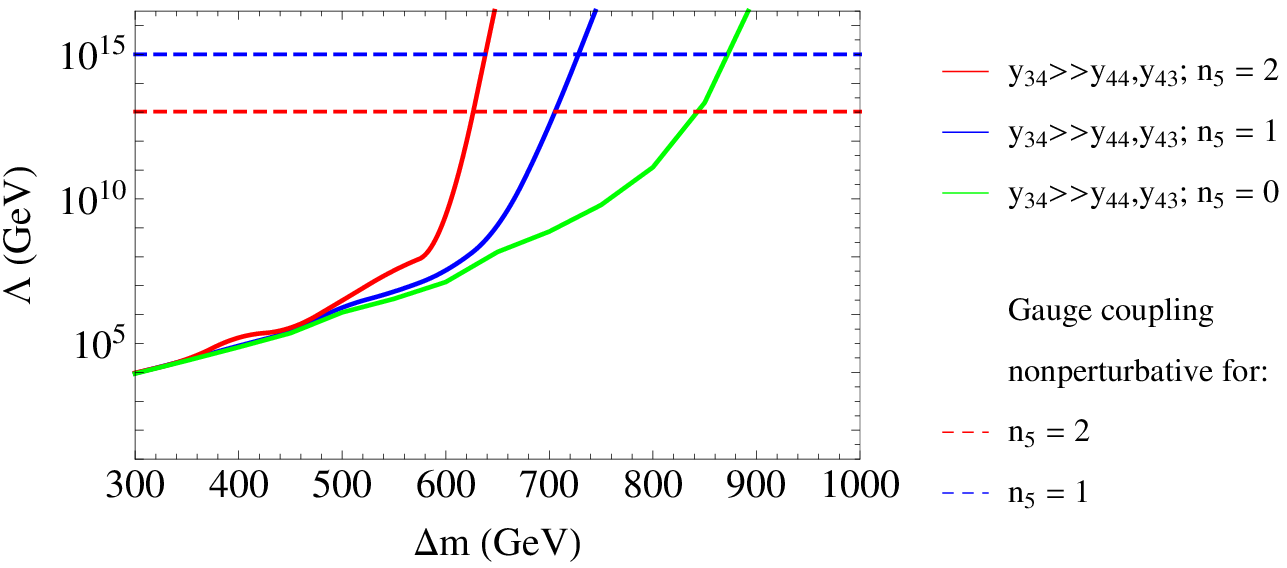}
\caption{We plot the scale $\Lambda$ where the $y_{33}$ required to get $m_h=125.5$ GeV hits a Landau pole, as a function of the soft mass $\Delta m$. We set $y_{34} \gg y_{44}, y_{43}$, $A=\Delta m$, $\mu_4 = 900$ GeV. Here the dotted lines indicate where the gauge couplings become non-perturbative for $n_5 = 2$ and $n_5=1$. They remain perturbative all the way to the GUT scale for $n_5=0$. \label{Fig:messcomp} }
\end{figure}

\indent From Figure \ref{Fig:yscales}, we see that large mixing can push $\Lambda$ above the GUT scale while retaining soft masses as low as $\sim 900$ GeV. The three different mixing scenarios give comparable results because these Yukawa couplings reinforce each other in their respective renormalization group evolution. In contrast, to push $\Lambda$ above $\sim10^{16}$ in the case with no mixing requires soft masses to be larger than 1.5 TeV.
\\
\indent From Figures \ref{Fig:Aterms34} and \ref{Fig:vecplots} it is clear that for a given soft mass, the implied Landau pole scale can also get pushed up by including larger $A$-terms or a smaller vector mass. For $A=\Delta m \sim 900$ GeV, $\Lambda$ can be pushed above the GUT scale. $\Delta m$ can be as low as 800 GeV, albeit in parts of parameter space with a Landau pole at $\sim 10^{10}$ GeV.
\\
\indent In the last point (4) above, we included one more parameter in our analysis, namely, the number $n_5$ of multiplets in the ${\bf 5 } + {\bf \bar{5}}$ representation of $SU(5)$ that are added to the model. These could correspond, for example in the minimal version of gauge-mediated supersymmetry breaking (GMSB), to messenger fields which don't couple to the Higgs and that communicate SUSY breaking from a hidden sector to the visible sector. This number does not affect the Yukawas necessary to obtain the Higgs mass but it contributes to the running of the gauge couplings, making them stronger in the ultraviolet.  And since the gauge couplings contribute negatively to the renormalization of the Yukawas, a larger ultraviolet gauge coupling slows the growth of the $y_{ij}$'s, pushing up the Landau pole. However, as we will see, to preserve perturbative gauge coupling unification we cannot add an arbitrary number of $n_5$ in addition to the vector-like  ${\bf 10 } + {\bf \bar{10}}$ of SU(5) necessary in our model. To verify perturbativity we used the one-loop beta functions presented in appendix \ref{app:BetaFunctionsGauge} and required $g_{\text{unif}}\lesssim3$. From Figure \ref{Fig:messcomp}, we see that the gauge couplings become non-perturbative around $10^{13}$ GeV for $n_5 = 2$ and $10^{15}$ GeV for $n_5 = 1$. They remain perturbative all the way to the GUT scale for $n_5=0$. Therefore, the Landau pole can still be pushed above the GUT scale if one sacrifices perturbativity at the scale of unification.

\section{Constraints}\label{Sec:Constraints}
In this section, we work out the constraints from Higgs production, measurements of the relevant Cabibbo-Kobayashi-Maskawa matrix element $V^{\text{CKM}}_{tb}$, the most recent mass bounds from direct searches for vector-like quarks at the LHC (with up to 19.5 $fb^{-1}$ of 8 TeV data from CMS \cite{CMS:2014} and 14.3 $fb^{-1}$ of 8 TeV data from the ATLAS detector) and constraints on the oblique parameters $S$ and $T$ \cite{SandToriginal} from electroweak precision measurements. We find that the oblique corrections and LHC direct searches place the dominant constraints on the total parameter space but that portions of the remaining parameter space available can still raise the Higgs mass to $\sim 125$ GeV while yielding new quarks discoverable at the LHC in the near future.

\subsection{Higgs Production}
The Higgs production rate at the LHC is dominated by the gluon fusion process and recent measurements can be used to put constraints on any model with new particles that get their mass through the Higgs. In the case where a chiral fourth generation is added to the SM, this leads to an increase of the Higgs production rate by gluon fusion by about a factor of nine over the SM rate, in contradiction with experiments. This is a result of the fact that the new quarks get all of their mass via coupling to the Higgs; no decoupling limit exists to ameliorate the situation.  However, in the case of a new generation of vector-like quarks the new quarks get their mass only partially through the Higgs, the remaining part coming from the vector-like mass parameter(s), here $\mu_4$. This opens the possibility that the new generation might contribute differently to Higgs production.  

One can see the dependence of the relevant amplitude on the parameters of the model as follows. We take the large tan$\beta$ limit throughout this discussion, though the procedure can be generalized in an obvious way. Consider an effective vertex coupling two gluons and a Higgs, which can be thought of as arising from a term in an effective Lagrangian with the form

\begin{equation}
\mathcal{L}_0 = g^* G_{\mu\nu} G^{\mu\nu} H,
\end{equation}
where $H \rightarrow h+v$ after electroweak symmetry breaking (EWSB) so that $\mathcal{L}_0 \rightarrow \mathcal{L}_1+\mathcal{L}_1'$, where

\begin{equation}
\mathcal{L}_1 = g^* G_{\mu\nu} G^{\mu\nu} h \quad,   \quad \mathcal{L}_1' = g^* G_{\mu\nu} G^{\mu\nu} v.
\end{equation}
The amplitude associated with the effective $ ggh $ vertex is simply the unknown $ g^*$. This is the same amplitude as for the $ \mathcal{L}_1'$ ``vertex," which can be interpreted as a correction to the gluon self-energy $\Pi^{gg} $.  In particular,  it is that part of the self-energy that comes from the coupling of particles in the loop to the Higgs vacuum expectation value (we consider only the one-loop correction). Rather than directly computing the effective $ggh$ coupling $g^*$ by summing all one-loop $ gg \rightarrow h $ diagrams, we can use the $ggv$ coupling to obtain $g^*$ from the well-known form of the gluon self-energy in a simple way.  For this we need consider all the contributions to the one-loop gluon self-energy, identify all the terms that include a factor of $v$, and sum the coefficients of $v$ from each term.  (Actually, what we need is just the sum, not individual coefficients.) Therefore to extract the information we want out of $\Pi^{gg}$, all we have to do is take a partial derivative with respect to $v$.  In equation form, $ g^* \sim \frac {\partial}{\partial v}  \left[\Pi^{gg}(v)\right]$, where $\Pi^{gg}$ is thought of as a function of $v$. 

The form of corrections to vector boson propagators is well known. Since the coupling for a non-Abelian gauge theory is universal,  all colored fermions in the loop contribute in the same way, i.e., the only difference between their contributions comes from the mass dependence. In particular, for a given quark running in the loop, one obtains a logarithmic dependence on its squared mass, $m_i^2$. This implies that 
\begin{equation}
\Pi^{gg} \supset c \sum_{i} \textnormal{log}(m_i^2),
\end{equation}
where $c$ is some constant and the sum is over $t, t_1', t_2'$. Now in the case under consideration all of the squared masses $m_i^2$ are the eigenvalues of the matrix $m_f^u m_f^{u \dagger}$ (as given in Appendix \ref{app:Spectrum}).  Since $\sum_{i} \textnormal{log} \left( m_i^2 \right) =  \textnormal{log} \left( \Pi_i m_i^2 \right) = \textnormal{log} \left[\textnormal{det} \left( m_f^u m_f^{u\dagger} \right) \right]$ and $\textnormal{det} \left( m_f^u m_f^{u\dagger} \right) = \textnormal{det}^2 \left( m_f^u  \right)$, the relevant terms in $\Pi^{gg}$ are given by
\begin{equation}
\Pi^{gg} \supset c \sum_{i} \textnormal{log} \left( m_i^2 \right) = c  \textnormal{ log} \left[\textnormal{det}^2 \left( m_f^u  \right) \right].
\end{equation}

Taking the partial derivative, 
\begin{equation}
A_{gg \rightarrow h} \propto \frac{\partial [ \text{log} (\det^2{m_f^u})]}{\partial v} = \frac{1}{\det^2{m_f^u}} \frac{\partial \det^2{m_f^u}}{\partial v}.
\end{equation}
In the special case $\bar{y}_{44}=0$, we have  $\text{det} (m_f^u) =v(y_{33}\mu^2_4\sin{\beta}),$
which (taking sin$\beta \approx 1$) is the same as in the SM aside from the factor of $\mu_4^2$, which cancels in the amplitude.  Thus $A_{gg \rightarrow h} \propto 2/v$, with no dependence on the $y_{ij}$'s or the vector-like mass parameter $\mu_4$, and there is no change from the well-known approximate SM amplitude.  We ignore contributions from the scalars, as these are suppressed. We note in passing that this expression has the right mass dimension for the $g^*$ mutiplying the dimension five operator in $\mathcal{L}_0$.

\subsection{$V^{\text{CKM}}_{tb}$}
The addition of the vector-like fourth generation will affect both the weak charged currents (CC) and the weak neutral currents (NC) at tree level. In particular, the $W^{\pm}$ gauge bosons now couple to both left-handed and right-handed particles. Furthermore, including mixing with the top sector will enrich the flavor structure of the model and induce flavor changing neutral currents (FCNCs) in the mass eigenstate basis. These FCNCs only involve third and fourth generation particles and are therefore fairly unconstrained. In appendix \ref{app:Couplings} we derive the triple and quartic gauge boson interaction terms with the quarks and squarks, as well as the interaction terms between the Higgs $h_o$ and quarks.
\\
\indent The rotation from gauge to mass eigenstates leads to generalized CKM matrices between the third generation, fourth generation, and it's mirror generation (which can be viewed as a ``fifth'' generation), which we denote by $K^{ab}_{\alpha}$ for quarks, and $\tilde{K}^{ab}_{\alpha}$ for squarks, with $a,b=u,\bar{u},d,\bar{d}$ and $\alpha=L,R$. These matrices will be present in every interaction term. Furthermore, they are not square matrices like in the MSSM because there are more up-type quarks than down-type quarks. \\
\indent The generalized CKM matrix $K^{ud}_L$ is a rectangular ($2\times3$) matrix (see appendix \ref{app:Couplings} for more details) in the mass basis $(t,t'_1,t'_2)$ for the (4-component) up-type quarks and $(b,b')$ for the down-type quarks. This matrix, being rectangular, is not unitary but satisfies the following equation:
\begin{align*}\label{eq:restriction}
K^{ud}_L (K^{ud}_L)^{\dagger} + K^{\bar{u}\bar{u}}_L (K^{\bar{u}\bar{u}}_L)^{\dagger} &= (V^{u\dagger}_L D^{ud}_L V^d_L)(V^{u\dagger}_L D^{ud}_L V^d_L)^{\dagger} +(V^{u\dagger}_L S^{\bar{u}\bar{u}\dagger}_L V^u_L)(V^{u\dagger}_L S^{\bar{u}\bar{u}\dagger}_L V^u_L)^{\dagger} \\
&= V^{u\dagger}_L(D^{ud}_L+S^{\bar{u}\bar{u}}_L)V^{u}_L \\
&= 1_{3\times3}
\end{align*}
where we have used the unitarity of the mixing matrices $V^{u}_L$ and $V^{d}_L$, and the fact that $D^{ud}_L (D^{ud}_L)^\dagger = D^{uu}_L$, $(S^{uu}_L)^{\dagger} S^{uu}_L = S^{\bar{u}\bar{u}}_L$ and $D^{uu}_L + S^{\bar{u}\bar{u}}_L = 1_{3\times3}$ (see appendix \ref{app:DsAndSs} for the explicit form of these matrices).\\
\indent The $(K^{ud}_L)_{11}$ entry predicted by our model should lie within the margin of error of the measured value of $V^{\text{CKM}}_{tb}$ (defined as the (3,3) entry of the ($3\times3$) matrix corresponding to the SM CKM matrix $V^{CKM}$). As usual, we neglect the mixing between the first two generations and the higher generations. When unitary of the SM $V^{CKM}$ is not assumed, $V^{\text{CKM}}_{tb}$ was recently measured by CMS \cite{CKMVtb}  to be $|V^{\text{CKM}}_{tb}| = 1.14 \pm 0.22$. We therefore require  $0.92 < (K^{ud}_L)_{11} <1.36$. After scanning over a large region of our relevant parameter space, we conclude that this restriction is always satisfied. Therefore, the constraints from the measured value of $V^{\text{CKM}}_{tb}$ are negligible. This is in agreement with the statements in \cite{Martin:2012dg}. 

\subsection{Mass Bounds from LHC Direct Searches}\label{subsection:Massbounds}
\begin{figure}
\includegraphics[width = 6.0 in]{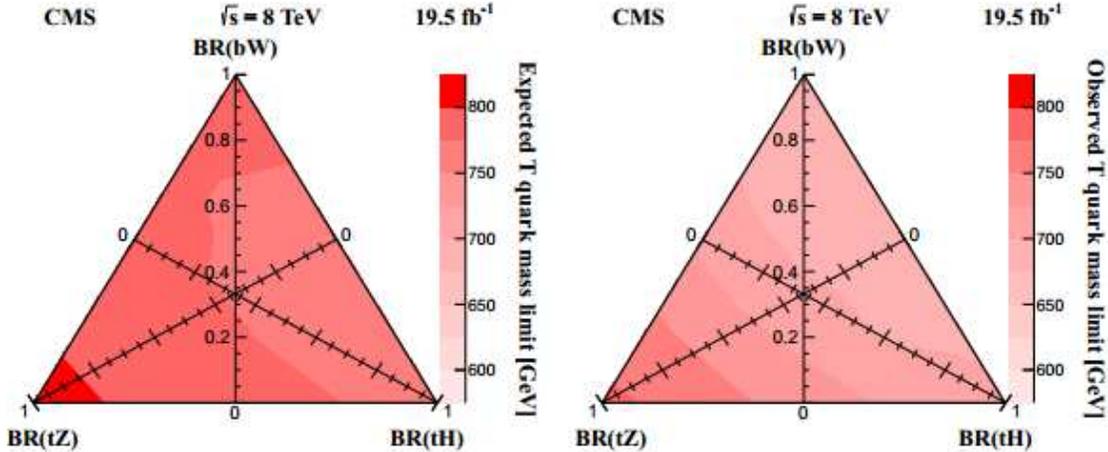}
\caption{Present status of heavy vector-like top searches with 19.5 fb$^{-1}$ of 8 TeV data with the CMS detector (Figure taken from \cite{CMS:2014}). A Branching-fraction triangle is shown with expected (left) and observed $95\%$ CL limits (right) on the mass. Every point in the triangle corresponds to a specific set of branching-fraction values subject to the constraint that all three add up to 1. \label{Fig:DirectConstrainsCMS} }
 \end{figure}

LHC direct searches \cite{Aaltonen:2008af, Chatrchyan:2012yea, ATLAS:2012aw, CMS:1209,CMS:2012ab, Aad:2011yn} are the most obvious source of constraints on the masses of the new vector-like quarks. The branching ratios (BRs) of the new quarks depend on the relative size of the relevant Yukawa, $W$ and $Z$ couplings.  Until fairly recently, many searches assumed 100 \% BR through one channel, particularly the $W b$ decay, and therefore had a large degree of model-dependence \cite{Geller:2012wx}. However, unlike these searches, ATLAS and CMS now can exclude vector-like quarks in a model independent way by considering general branching ratio scenarios in their data analysis \cite{CMS:2014}.\\
\indent At the LHC, the $t'$ (or $b'$) can be either pair produced or singly produced. Typically, the pair produced initial state has a large cross section, however, as shown in \cite{VLQ:2013} it is possible that single production of the heavy quark via the exchange of a $t$-channel $W$ have a larger cross section than $t' t'$. This opens new decay chains such as $t' b j \rightarrow ht b j \rightarrow bb Wb b j$. In Table \ref{Tab:SingleProd} and Table \ref{Tab:PairProd} we list possible event topologies that could arise at the LHC. For the final states, we see that there may be as many as six $b$ jets, or if the Higgs decays via the less common $WW^*$ channel then there may be as many as six $W$ bosons. Finally, we note that $t' b j \rightarrow Wb b j$ and $t' t' \rightarrow Wb Wb$ present two of the best routes to discovery since $m_{Wb}$ would reconstruct to $m_{t'}$ and the signals are relatively clean.
\begin{table}
\renewcommand*\arraystretch{1.0}
\begin{tabular}{|c|c|c||c|c|c|}
  \hline
  \textbf{Initial}  & \textbf{Intermediate} & \textbf{Final} & \textbf{Initial} & \textbf{Intermediate} & \textbf{Final} \\\hline\hline
  $t'$ & $ht$ & $bbWb$ & $b'$ & $hb$ & $bbb$ \\\hline
  $t'$ & $Zt$ & $ffWb$ & $b'$ & $Zb$ & $ffb$ \\\hline
  $t'$ & $Wb$ & $Wb$ & $b'$ & $Wt$  & $WWb$ \\\hline
  $t't$ & $htt$ & $bbWbWb$ & $b'b$ & $hb$ & $bbbb$ \\\hline
  $t't$ & $Ztt$ & $ffWbWb$ & $b'b$ & $Zb$ & $ffbb$ \\\hline
  $t't$ & $Wbt$ & $WbWb$ & $b'b$ & $Wtb$ & $WWbb$ \\\hline
  $t'bj$ & $htbj$ & $bbWbbj$ & $b'tj$ & $hbWbj$ & $bbbWbj$ \\\hline
  $t'bj$ & $Ztbj$ & $ffWbbj$ & $b'tj$ & $ZbWbj$ & $ffbWbj$ \\\hline
  $t'bj$ & $Wbbj$ & $Wbbj$ & $b'tj$ & $WtWbj$ & $WWbWbj$ \\\hline
\end{tabular}
\caption{ Possible event topologies that could arise at the LHC with initial states involving only one single $t'$ or $b'$. $f$ denotes any fermion, $(f=q,l)$}
\label{Tab:SingleProd}
\end{table}
\begin{table}
\renewcommand*\arraystretch{1.0}
\begin{tabular}{|c|c|c||c|c|c|}
  \hline
  \textbf{Initial}  & \textbf{Intermediate} & \textbf{Final} & \textbf{Initial} & \textbf{Intermediate} & \textbf{Final} \\\hline\hline
  $t't'$ & $htht$ & $bbWbbbWb$ & $b'b'$ & $hbhb$ & $bbbbbb$ \\\hline
  $t't'$ & $htZt$ & $bbWbffWb$ & $b'b'$ & $hbZb$ & $bbbffb$ \\\hline
  $t't'$ & $htWb$ & $bbWbWb$ & $b'b'$ & $hbWt$  & $bbWWb$ \\\hline
  $t't'$ & $ZtZt$ & $ffWbffWb$ & $b'b'$ & $ZbZb$ & $ffbffb$ \\\hline
  $t't'$ & $ZtWb$ & $ffWbWb$ & $b'b'$ & $ZbWt$ & $ffbWWb$ \\\hline
  $t't'$ & $WbWb$ & $WbWb$ & $b'b'$ & $WtWt$ & $WWbWWb$ \\\hline
\end{tabular}
\caption{ Possible event topologies that could arise at the LHC with initial states involving a pair produced $t'$ or $b'$. $f$ denotes any fermion, $(f=q,l)$}
\label{Tab:PairProd}
\end{table}
\\
\indent 
The most recent search done by CMS is the first search to consider all the three final states, and puts the most stringent constraints to date on the existence of a heavy vector-like top quark. Assuming that the heavy vector-like top quark decays exclusively into $bW$, $tZ$, and $tH$, CMS has set lower limits for its mass between 687 and 782 GeV for all possible branching fractions into these three final states assuming strong production. Their results are summarized in Figure \ref{Fig:DirectConstrainsCMS} (taken from \cite{CMS:2014}).
\\
\indent For ATLAS, the high multiplicity of jets has recently been used in the search for vector-like quarks, yielding the mass bound on the $t'$ consistent with CMS \cite{ATLASconf:2013-018}.
Therefore, requiring the vector-like mass parameter $\mu_4\gtrsim700$ ensures that the physical masses of the new heavy quarks are above the lower bounds excluded by the LHC.

\subsection{Electroweak Precision Observables}\label{subsection:EPO}
\begin{figure}
\includegraphics[width = 4.0 in]{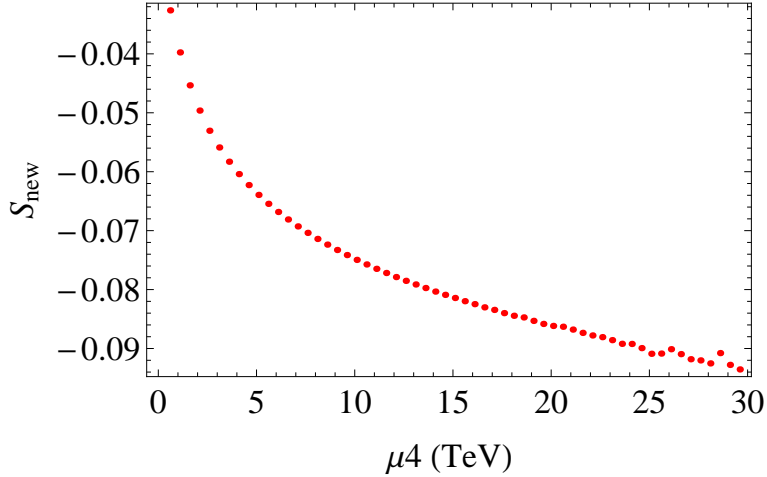}
 \caption{ $S_{new}$ versus $\mu_4$ for $y_{34}=0.6$ and $y_{44}=y_{43}=0$. $S_{new}$ remains small as $\mu_4\rightarrow\infty$. \label{Fig:Smudependence}}
\end{figure}

\begin{figure}
\includegraphics[width = 4.0 in]{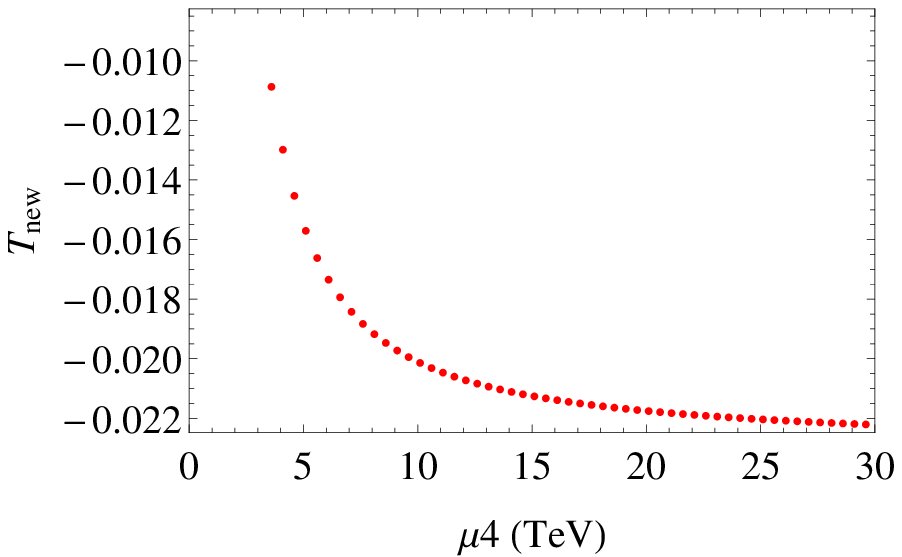}
 \caption{$T_{new}$ versus $\mu_4$ for $y_{34}=0.6$ and $y_{44}=y_{43}=0$. $T_{new}$ remains small as $\mu_4\rightarrow\infty$. \label{Fig:Tmudependence}}
\end{figure}

\begin{figure}
\includegraphics[width = 4.2 in]{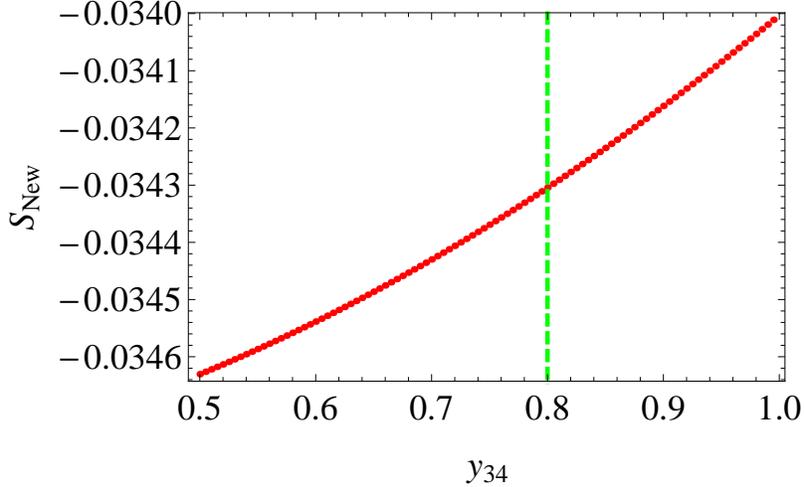}
 \caption{ $S_{new}$ versus $y_{34}$ for the benchmark scenario $y_{34}>>y_{44},y_{43}$, $\mu_4=900$ GeV, $A=\Delta m=800$ GeV. $S_{new}$ remains small in this region. As $y_{34}$ increases from 0.5 to 1, $S_{new}$ increases by a negligible amount of the order of $10^{-4}$. The region $y_{34}\gtrsim0.8$ to the right of the dashed line is disfavored by EWPM due to the $T$ parameter (see Figure \ref{Fig:TYukdependence}).  \label{Fig:SYukdependence}}
\end{figure}

\begin{figure}
\includegraphics[width = 3.9 in]{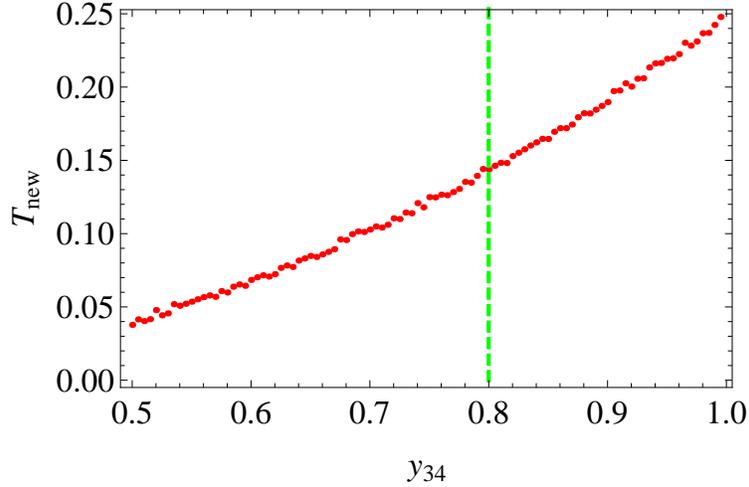}
 \caption{ $T_{new}$ versus $y_{34}$ for the benchmark scenario $y_{34}>>y_{44},y_{43}$, $\mu_4=900$ GeV, $A=\Delta m=800$ GeV. As $y_{34}$ increases from 0.5 to 1, $T_{new}$ increases from $\sim0.05$ to $\sim 0.25$. The region $y_{34}\gtrsim0.8$ to the right of the dashed line is disfavored by EWPM as can be seen in Figure \ref{Fig:MasterScan}. \label{Fig:TYukdependence}}
\end{figure}

\begin{figure}
\includegraphics[width = 5.75 in]{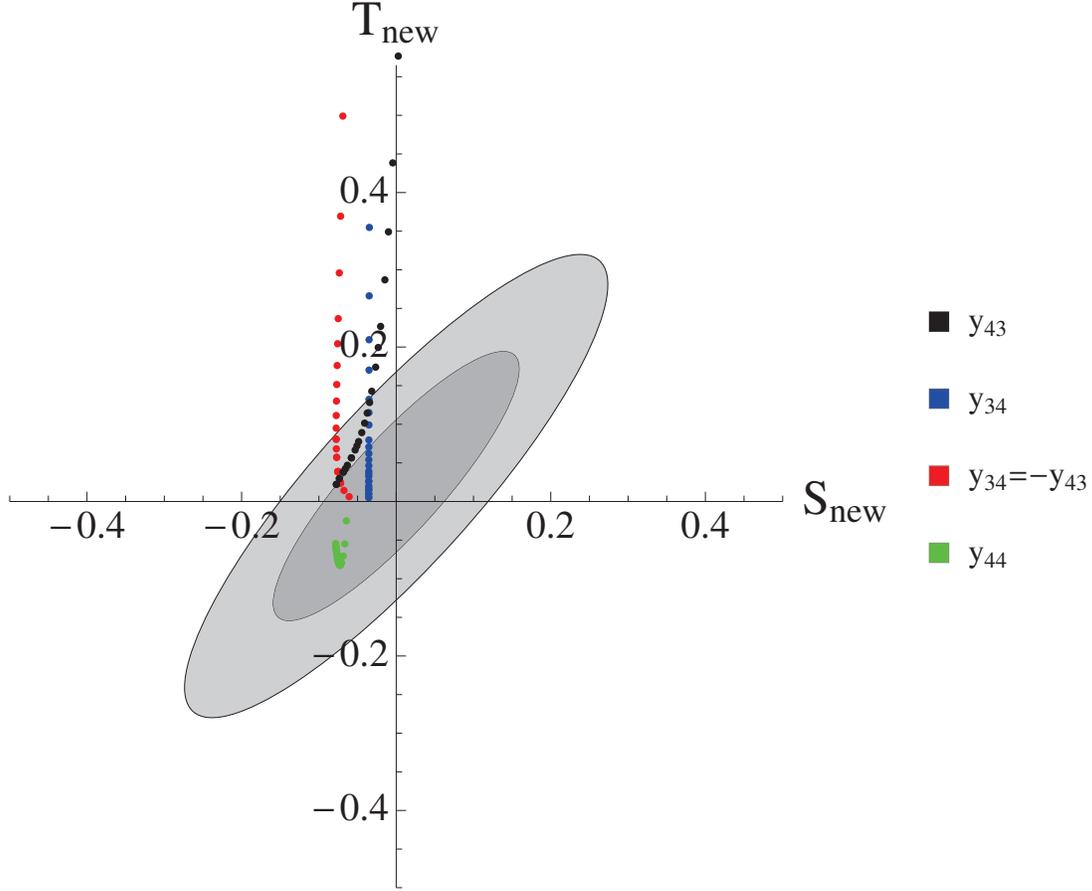}
 \caption{ We calculate $S_{new}$ and $T_{new}$ for each of the benchmark scenarios: $y_{34}>>y_{43}, y_{44}$; $y_{43}>>y_{34}, y_{44}$; and $y_{34}=-y_{43}>>y_{44}$.  Within each scenario $\mu_4 = 900$ GeV, $A=600$ GeV, and we vary $\Delta m$ from $300$ to $1500$ GeV.  Each of these points satisfies current mass bounds (see section \ref{subsection:Massbounds}) and gives a Higgs mass $m_h=125.5 \pm .5$ GeV while yielding new quarks discoverable at the LHC. The points corresponding to very low $\Delta m$ and larger Yukawas lie farthest from the best fit, with the agreement improving as $\Delta m$ grows and the Yukawas decrease. For many of these points the net effect from the new sector falls within the 95\% or 68\% confidence limits on the electroweak observables. The experimental best fit corresponds to the center of the ellipses, at $(0.00, 0.02)$ \cite{EWPD:2013PDG}. The light (dark) grey ellipse denote the 95\% (65\%) CL on the EW observables. The origin is defined to be the Standard Model prediction with a 125 GeV Higgs. In concert with the results of section \ref{Sec:HiggsMass}, precision electroweak observables permit sufficiently large Yukawa mixing to obtain a Higgs mass $\sim 125$ GeV with soft terms below a TeV.\label{Fig:MasterScan} }
\end{figure}

\begin{figure}
\includegraphics[width = 5.75 in]{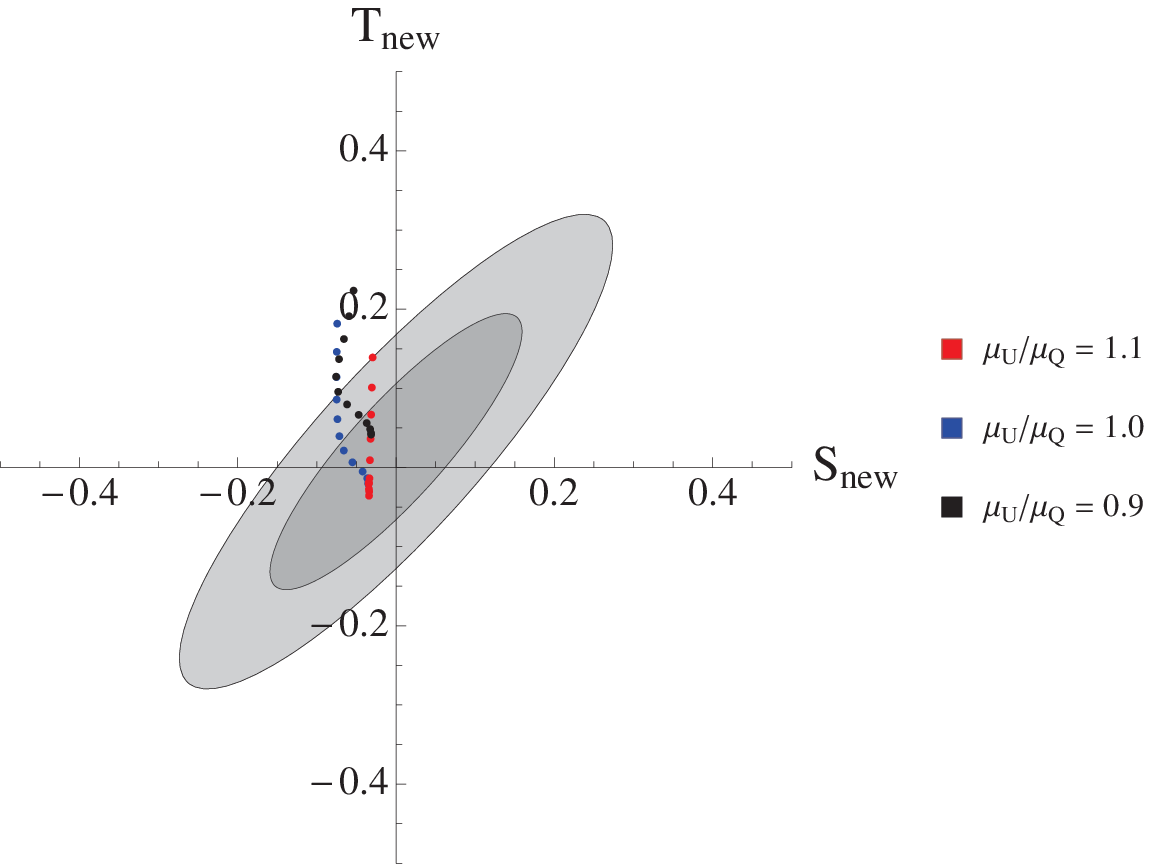}
 \caption{  We plot the $S_{new},T_{new}$ for ratios $\mu_U/\mu_Q = 0.9, 1.0, 1.1$, and Yukawa values $y_{34}=-y_{43}$ ranging from $0.01$ to $0.56$ in steps of $0.05$.  Each of these points satisfies current mass bounds (see section \ref{subsection:Massbounds}) and gives a Higgs mass $m_h=125.5 \pm .5$ GeV while yielding new quarks discoverable at the LHC. The points corresponding to very low $\Delta m$ and larger Yukawas lie farthest from the best fit, with the agreement improving as $\Delta m$ grows and the Yukawas decrease. For many of these points the net effect from the new sector falls within the 95\% or 68\% confidence limits on the electroweak observables. The experimental best fit corresponds to the center of the ellipses, at $(0.00, 0.02)$ \cite{EWPD:2013PDG}. The light (dark) grey ellipse denote the 95\% (65\%) CL on the EW observables. The origin is defined to be the Standard Model prediction with a 125 GeV Higgs. In concert with the results of section \ref{Sec:HiggsMass}, precision electroweak observables permit sufficiently large Yukawa mixing to obtain a Higgs mass $\sim 125$ GeV with soft terms below a TeV.\label{Fig:RScatterPlot} }
\end{figure}

We now study the total contribution of the new generation to the electroweak oblique parameters $S$ and $T$. In appendix \ref{app:Couplings}, we work out the interaction terms between the new particles and the electroweak gauge bosons in the mass basis Lagrangian, as these are needed to derive the necessary Feynman rules to calculate the self energy loops in the definitions of $S$ and $T$. The relevant interaction terms are of the form $Wff$, $Zff$, $Aff$ and for quarks, and $W\tilde{f}\tilde{f}$, $Z\tilde{f}\tilde{f}$, $A\tilde{f}\tilde{f}$, $WW\tilde{f}\tilde{f}$, $ZZ\tilde{f}\tilde{f}$, $AA\tilde{f}\tilde{f}$ and $ZA\tilde{f}\tilde{f}$ for squarks. In appendix \ref{app:SandT} we calculate the contributions to the oblique parameters from both fermions ($T_f$, $S_f$) and scalars ($T_s$, $S_s$). We note that in the full decoupling limit, $\mu_4\rightarrow\infty$ and $y_{ij}\rightarrow 0$ we recover SM values.
\\
\indent To get the total contribution of the new sector, we define $T_{new}=T_f+T_s - T_{SM}$ and $S_{new}=S_f+S_s - S_{SM}$. The values $T_{SM}\approx 1.22$ and $S_{SM}\approx -0.08$ were calculated to account for the top sector alone. In general, we find that $T_s<<T_f$ and $S_f \approx S_s$.
\\
\indent The $\mu_4$ dependences of $S_{new}$ and $T_{new}$ are shown in Figures \ref{Fig:Smudependence} and \ref{Fig:Tmudependence}, respectively, for the benchmark scenario $y_{34}=0.6$ and $y_{44}=y_{43}=0$ with the Yukawa values kept fixed. As a sanity check, we see that for a large range of $\mu_4$, the values of $S$ and $T$ remain very small.
\\
\indent The dependences of $S_{new}$ and $T_{new}$ on the mixing Yukawa couplings are shown in Figures and \ref{Fig:SYukdependence} and \ref{Fig:TYukdependence}, respectively, for the benchmark scenario $y_{34}>>y_{44},y_{43}$ with $\mu_4=900$ GeV kept fixed and $A=\Delta m=800$ GeV. As $y_{34}$ increases from 0.5 to 1, $S_{new}$ increases by a negligible amount of the order of $10^{-4}$. However, $T_{new}$ increases by $\sim 0.25$. For $T\gtrsim 0.15$, there is tension with the EWPM fit (as can be seen in Figure \ref{Fig:MasterScan}) and therefore the maximum allowed value for $y_{34}$ in this case is $\sim0.8$.
\\
\indent To get a more general picture, we scanned over a wide range of the parameter space from the new sector consistent with the mass bounds from the LHC (see section \ref{subsection:Massbounds}). We varied the relevant $y_{ij}$'s, $\mu_4$, and $\Delta m$ but kept the $A$-terms fixed at 800 GeV. The results are presented in Figure \ref{Fig:MasterScan}. We see that $-0.1\lesssim S_{new} \lesssim 0$, while $T_{new}$ can be positive or negative. The positive contributions of  $T_{new}$ can be large enough to be in tension with EWPD.  Nevertheless, from Figure \ref{Fig:MasterScan} it is clear that with vector masses $\mu_4 \gtrsim 900$ GeV a large set of our parameter space of interest falls within the 95\% and 68\% confidence limits on the electroweak observables.
\\
\indent Furthermore, while taking $\mu_U/\mu_Q = 1$ is a natural simplification, in general this condition does not hold.  Indeed, if the vector masses are taken to be equal at some high SUSY-breaking scale, then differences in the beta functions will result in unequal vector masses at the weak scale.  We therefore probed the effect of varying this ratio while keeping the sum of the masses constant. The ratio is less constrained for smaller mixing Yukawas, with  $2.3 \gtrsim \mu_U/\mu_Q \gtrsim 0.85$ allowed by EWPM for $y_{34}=-y_{43}=0.1$ and $\mu_Q+\mu_U=1800$ GeV, while for large $y_{34}=-y_{43}$ we find $1.2 \gtrsim \mu_U/\mu_Q \gtrsim 0.9$.  On the other hand, there are scenarios in which the effects from a non-unity ratio value counteract the effects from large mixing Yukawas.  For example, with  $\mu_U/\mu_Q = 1.1$ it was found that $y_{34}=-y_{43}$ can be as large as 0.56 and still fall within the 95\% confidence limits on EWPD, up from 0.43 for a ratio of one.  Since EWPM give the most significant constraints on the $y_{ij}$'s, we see by referring to Figure \ref{Fig:yvalues} that soft masses $\lesssim 800$ GeV are then the minimum required for the $y_{34}=-y_{43}$ case, rather than the $\sim 1000$ GeV it requires when the ratio is one (the $y_{ij}$'s needed to give the desired Higgs mass have negligible dependence on the value of the ratio). In Figure \ref{Fig:RScatterPlot} we plot the $S_{new},T_{new}$ for ratios $\mu_U/\mu_Q = 0.9, 1.0, 1.1$, and Yukawa values $y_{34}=-y_{43}$ ranging from $0.01$ to $0.56$ in steps of $0.05$.
\\
\indent  We conclude that in concert with the results of section \ref{Sec:HiggsMass}, precision electroweak observables permit sufficiently large Yukawa mixing to obtain a Higgs mass $\sim 125$ GeV with  soft parameters below a TeV while yielding new quarks discoverable at the LHC.

\section{Conclusions}\label{sub:conclusion}
In this paper we studied the effects of sizeable mixing Yukawa terms between the top sector and a vector-like quark generation. We computed the energy scale of the Landau pole induced by the top Yukawa for various scenarios. We also discussed the LHC phenomenology and the consequences of including top mixing effects on final state event topologies.
\\
\indent We found that sizeable mixing Yukawa couplings ($y_{34}$ and $y_{43}$) in the superpotential require an increase of the value of the top Yukawa coupling by at most $\sim6\%$ to produce the observed top mass. Since loop corrections to $m_h$ go as $y_{top}^4$, mixing will increase the predicted value of the physical Higgs mass, a point not previously emphasized in the literature. This high sensitivity to the top Yukawa is in contrast with the weaker logarithmic dependence on top squark masses.
\\
\indent The mixing Yukawas necessary to achieve a given Higgs mass are smaller when $ |y_{34}| \sim |y_{43}|$ than when one of these couplings dominates the other, and  if one allows $\mu_U/\mu_Q \neq 1$  then the lowest soft masses ($\Delta m \sim 750$ GeV) can be accommodated for this case.  However, under the restriction  $\mu_U/\mu_Q=1$, then the lowest possible value of $\Delta m$ consistent with EWPM is $\Delta m\sim 800$ GeV, which occurs when $y_{34}\sim0.8$ and $y_{43}=y_{44}=0$ (see Figure \ref{Fig:yvalues}).
\\
\indent Moreover, mixing can significantly raise the Higgs mass while retaining perturbativity to much higher scales than possible with only the self coupling $y_{44}$ of the fourth generation (see Figure \ref{Fig:yscales}). For $A$-terms and soft masses around 900 GeV, the top Yukawa Landau pole can be pushed above the GUT scale. For $\mu_Q=\mu_U$, soft masses can be as low as 800 GeV and still generate a Higgs mass of 125 GeV, albeit in parts of parameter space with a Landau pole at $\sim 10^{10}$ GeV. Smaller supersymmetry-breaking terms suffice if one sacrifices perturbativity at the unification scale by adding fields in a $\mathbf{5}$+$\mathbf{\bar{5}}$ (see Figure \ref{Fig:messcomp}).
\\
\indent We studied the constraints from electroweak precision measurements, the measurements of $V^{\text{CKM}}_{tb}$, Higgs production, and the most recent mass bounds from direct searches for vector-like quarks at the LHC. We found that the oblique corrections and LHC direct searches give the dominant constraints. With vector masses $\mu_4 \gtrsim900$ GeV and soft scalar masses $\Delta m\gtrsim 800$ GeV, the net effect from the new sector falls within the 95\% confidence limits on the electroweak observables.
\\
\indent We conclude that there is a large parameter space available for a supersymmetric model with a vector-like fourth generation that passes all tests from previous experimental analyses with sufficiently large Yukawa mixing to obtain a Higgs mass $\sim 125$ GeV, while yielding new quarks discoverable at the LHC. These models have a soft SUSY breaking scale that remains moderate and can therefore address the little hierarchy problem.
\\
\indent Finally, we refer to the appendix for details about the particle spectrum, the derivation of the mass matrices in the model and the calculation of the oblique parameters. In addition, we give the explicit form of all of the matrices needed to write the interaction Lagrangian. These include generalized CKM matrices, couplings matrices and projection matrices. We also list the beta functions used in the study of Landau poles and perturbativity, as well as loop functions used in the calculation of the oblique parameters.

\appendix
\section{The Physical Spectrum and Mass Matrices}\label{app:Spectrum}
After the $SU(2)_L \times U(1)_Y$ gauge symmetry is broken, Yukawa terms in the superpotential (equation \ref{Eq:Superpotential}), soft terms, $F$ terms, and $D$ terms lead to the following fermion mass matrices:

\begin{equation*}
M_f^u= \left(
\begin{array}{cccccc}
 0 & m_f^u \\
m_f^{u\dagger} & 0
\end{array}
\right),
\qquad\text{with}\qquad
m_f^u \equiv \left(
\begin{array}{cccccc}
y_{33}v_u & y_{34}v_u & 0 \\
y_{43}v_u & y_{44}v_u & \mu_Q \\
0 & \mu_U & 0 \\
\end{array}
\right),
\end{equation*}
\\
\begin{equation*}
M_f^d \equiv \left(
\begin{array}{cccccc}
0 & m_f^d \\
m_f^{d\dagger} & 0 \\
\end{array}
\right),
\qquad\text{with}\qquad
m_f^d\equiv\left(
\begin{array}{cccccc}
m_{\text{bot}} & 0 \\
0 & \mu_Q \\
\end{array}
\right),
\end{equation*}
and the scalar squared mass matrices:
\begin{align*}
\small
(M_s^u)^2&= (M_f^u)^2
+ \left(
\begin{array}{cccccc}
Y_{u_{3}} & 0 & 0 & -y_{33}v_u X_u & -y_{34}v_u X_u  & 0 \\
0 & \mu_Q^2 + Y_{u_{4}} & 0 & -y_{43}v_u X_u  & -y_{44}v_u X_u & B\mu \\
0 & 0 & \mu_U^2 + Y_{\bar{u}_{4}} & 0 & B\mu & 0 \\
-y_{33}v_u X_u & -y_{43}v_u X_u & 0 & Y_{u^c_{3}} & 0 & 0 \\
-y_{34}v_u X_u & -y_{44}v_u X_u & B\mu & 0 & \mu_U^2 + Y_{u^c_{4}} & 0 \\
0 & B\mu & 0 & 0 & 0 & \mu_Q^2 + Y_{\bar{u}^{c}_{4}} \\
\end{array}
\right),
\end{align*}
\\
\begin{equation*}
(M_s^d)^2=(M_f^d)^2+\left(
\begin{array}{cccccc}
 Y_{d_{3}} & 0 & -m_{\text{bot}}X_d & 0 \\
 0 & \mu_Q^2 + Y_{d_{4}} & 0 & B\mu \\
-m_{\text{bot}}X_d & 0 & Y_{d^c_{3}} & 0 \\
 0 & B\mu & 0 & \mu_Q^2 + Y_{\bar{d}^{c}_{4}} \\
\end{array}
\right).
\end{equation*}
\\
\indent Here, $v_u=v\sin\beta$, with $v\approx174$ GeV, and $m_{\text{bot}}\approx4.2$ GeV is the mass of the bottom quark. $X_u=A + \mu \cot\beta$ and $X_d=A + \mu \tan\beta$. Along the diagonal, $Y_q\equiv \Delta m^2 + D_a$, where the $D$-term contribution is $D_a=(T^3_a - Q_a\sin^2\theta_w)\cos(2\beta)m^2_Z$ for each quark field $a$, $T^3$ is the third component of weak isospin, $Q_a$ is the electric charge, and $\theta_w$ is the weak mixing angle. We take all parameters to be real. With the mass matrices defined as above, the relevant mass Lagrangian (after EWSB) in the gauge eigenstate basis can be written as:
\begin{equation}\label{eqn:MassLag}
-\mathcal{L}_m= (f_L^{uT} m_f^u f_R^u + f_L^{dT} m_f^d f_R^d + \text{h.c}) + \tilde{f}^{u\dagger} (M_s^u)^2 \tilde{f}^u + \tilde{f}^{d\dagger} (M_s^d)^2 \tilde{f}^d
\end{equation}
where the basis is:
\begin{align}\label{eq:fs}
\nonumber f_L^u&= (u_{3}, u_{4}, \bar{u}_{4})^T \\
\nonumber f_R^u&= (u^c_{3}, u^c_{4}, \bar{u}^c_{4})^T \\
\nonumber f_L^d&= (d_3, d_{4})^T \\
f_R^d&= (d^c_3, \bar{d}^c_{4})^T \\
\nonumber \tilde{f}^u&= (\tilde{u}_3, \tilde{u}_{4}, \tilde{\bar{u}}_{4}, \tilde{u}^c_{3}, \tilde{u}^c_{4}, \nonumber \tilde{\bar{u}}^c_{4})^T \\
\nonumber \tilde{f}^d&= (\tilde{d}_3, \tilde{d}_{4}, \tilde{d}^c_{3}, \tilde{\bar{d}}^c_{4})^T.
\end{align}
The physical masses of the fermions are obtained by bi-diagonalizing the fermion mass matrices using the singular value decomposition:
\begin{align*}
  m^u_D &= V_L^{u\dagger} m_f^u V^u_R \\
  m^d_D &= V_L^{d\dagger} m_f^d V^d_R
\end{align*}
where $V^{u,d}_L$ and $V^{u,d}_R$ are unitary matrices and the $m^{u,d}_D$ matrices are diagonal. The diagonal entries of $m_f^u$ ($m_f^d$) correspond to the physical masses of the top (bottom) and the new non-MSSM quarks $t'_{1,2}$ ($b'$). Similarly, the scalar squared matrices are diagonalized by the unitary matrices $W^{u,d}$ as:
\begin{align*}
  (\tilde{M}_D^u)^2 &= W^{u\dagger} (M_s^u)^2 W^u \\
  (\tilde{M}_D^d)^2 &= W^{d\dagger} (M_s^d)^2 W^d,
\end{align*}
where the $(\tilde{M}^{u,d}_D)^2$ matrices are diagonal. The positive square roots of $(\tilde{M}_D^u)^2$ (and $(\tilde{M}_D^d)^2$) correspond to the physical masses of the top squarks (bottom squarks) and the new non-MSSM squarks $\tilde{t}_{1,2}$, $\tilde{t}'_{1,2,3,4}$ ($\tilde{b}_{1,2}$, $\tilde{b}'_{1,2}$). To obtain a Lagrangian in the mass eigenstate basis, we rotate the gauge eigenstates by left-multiplying the vectors $f^{u,d}$ and $\tilde{f}^{u,d}$ in equation \ref{eq:fs} by the corresponding mixing matrices $V^{u,d\dagger}_{L,R}$ and $W^{u,d\dagger}$, respectively. We denote the mass eigenstate basis with a hat, $\hat{f}^{u,d}_{L,R}=V_{L,R}^{u,d\dagger}f^{u,d}_{L,R}$ and $\hat{\tilde{f}}^{u,d}=W^{u,d\dagger}\tilde{f}^{u,d}$. A typical particle spectrum is shown in Table \ref{Tab:table2} for $\mu_4=900$ GeV.

\begin{table}
\renewcommand*\arraystretch{1.0}
\begin{tabular}{|c|c|c|c|}
   \hline
  \textbf{ Mass (GeV)} &  \textbf{Scenario 1} &  \textbf{Scenario 2} &  \textbf{Scenario 3} \\\hline\hline
    $m_{t'_1}$& 909 & 900 & 900 \\\hline
    $m_{t'_2}$& 913 & 911 & 900 \\\hline
    $m_{b'}$& 900 & 900 & 900 \\\hline
    $\tilde{m}_{\tilde{t}_1}$& 814 & 818 & 821 \\\hline
    $\tilde{m}_{\tilde{t}_2}$& 982 & 991 & 1000 \\\hline
    $\tilde{m}_{\tilde{t}'_1}$& 1275 & 1271 & 1271 \\\hline
    $\tilde{m}_{\tilde{t}'_2}$& 1276 & 1273 & 1272 \\\hline
    $\tilde{m}_{\tilde{t}'_3}$& 1287 & 1275 & 1273 \\\hline
    $\tilde{m}_{\tilde{t}'_4}$& 1300 & 1294 & 1274 \\\hline
    $\tilde{m}_{\tilde{b}_1}$& 860 & 860 & 860 \\\hline
    $\tilde{m}_{\tilde{b}_2}$& 940 & 940 & 940 \\\hline
    $\tilde{m}_{\tilde{b}'_1}$& 1271 & 1271 & 1271  \\\hline
    $\tilde{m}_{\tilde{b}'_2}$& 1274 & 1275 & 1274 \\\hline
\end{tabular}
\caption{A typical particle spectrum for the three different benchmark scenarios: 1) $y_{34}=-y_{43}=0.8$ and $y_{44}=0$; 2) $y_{34}=0.8$ and $y_{43},y_{44}=0$; 3) $y_{44}=0.8$ and $y_{34}, y_{43}=0$. The scenario $y_{43}=0.8$ and $y_{34},y_{44}=0$ gives the same masses as scenario 2) and we therefore omit it. We set $A=\Delta m = \mu_4 = 900$ GeV. As we can see, mixing doesn't change drastically the mass spectrum.}
\label{Tab:table2}
\end{table}

\section{The Interaction Lagrangian}\label{app:Couplings}
The rotation from gauge to mass eigenstates leads to generalized CKM matrices between the third and fourth generation, which we denote by $K^{ab}_{\alpha}$ for quarks, and $\tilde{K}^{ab}_{\alpha}$ for squarks, with $a,b=u,\bar{u},d,\bar{d}$ and $\alpha=L,R$. These matrices will be present in every interaction term. Furthermore, they are not square matrices like in the MSSM because there are more up-type quarks (squarks) than down-type quarks (squarks). Their general form is $K^{ab}_{\alpha}=V^{a\dagger}_{\alpha}D^{ab}_{\alpha}V^{b}_{\alpha}$ or $K^{ab}_{\alpha}=V^{a\dagger}_{\alpha}S^{ab}_{\alpha}V^{b}_{\alpha}$, and $\tilde{K}^{ab}_{\alpha}=W^{\dagger}\tilde{D}^{ab}_{\alpha}W$ or $\tilde{K}^{ab}_{\alpha}=W^{\dagger}\tilde{S}^{ab}_{\alpha}W$. The projection matrices, $D^{ab}_{\alpha}$ and $\tilde{D}^{ab}_{\alpha}$ ($S^{ab}_{\alpha}$ and $\tilde{S}^{ab}_{\alpha}$) select the appropriate doublet (singlet) field component of $f^a$ and $\tilde{f^a}$, respectively, before rotating to the mass basis. We note that, in general, $K^{aa}_{\alpha}=K^{ab}_{\alpha}K^{ab\dagger}_{\alpha}$, so we can construct all of the generalized CKM matrices from all the possible products of $K^{ab}_{\alpha}$ and $K^{ab\dagger}_{\alpha}$. It is therefore the non-unitarity and off-diagonal entries of $K^{ab}_{\alpha}$ that leads to FCNC's. $K^{ab}_{\alpha}$  and $\tilde{K}^{ab}_{\alpha}$ depend on the flavor and chirality of the particles involved in the interaction, and on the parameters of the model (e.g. $\mu_4$,the $y_{ij}$'s) which are present in the corresponding mixing matrices $V^a_{\alpha}$ and $W^a$.
\\
\indent In Tables \ref{Tab:CKMmatricesF} and \ref{Tab:CKMmatricesS}, we give the form of all these generalized CKM matrices and write down the corresponding interaction term coupling the vector bosons $V_{\mu}=W_{\mu},Z_{\mu}, A_{\mu}$ to the quarks or squarks, in the mass basis. The matrices $D^{ab}_{\alpha}$, $\tilde{D}^{ab}_{\alpha}$, $S^{ab}_{\alpha}$ and $\tilde{S}^{ab}_{\alpha}$ are listed in appendix \ref{app:DsAndSs}, and the mixing matrices $V^a_{\alpha}$ and $W^a$ were calculated numerically and depend on the parameters of the model.

\begin{table}
\renewcommand*\arraystretch{1.0}
\begin{tabular}{|c|c|}
   \hline
   $\mathbf{V_{\mu}\hat{f}^{a\dagger}_{\alpha}K^{ab}_{\alpha}\bar{\sigma}^{\mu}\hat{f}^b_{\alpha}}$ & $\mathbf{K^{ab}_{\alpha}}$\\\hline\hline
   $W_{\mu}^{+}\hat{f}^{u\dagger}_LK^{ud}_L\bar{\sigma}^{\mu}\hat{f}^d_L$ & $V^{u\dagger}_LD^{ud}_L V^d_L$\\\hline
   $W_{\mu}^{+}\hat{f}^{d\dagger}_RK^{\bar{u}\bar{d}\dagger}_R\bar{\sigma}^{\mu}\hat{f}^u_R$ & $V^{u\dagger}_RD^{\bar{u}\bar{d}}_R V^d_R$\\\hline
   $Z_{\mu}^{0}\hat{f}^{u\dagger}_LK^{uu}_L\bar{\sigma}^{\mu}\hat{f}^u_L$ & $V^{u\dagger}_LD^{uu}_L V^u_L$\\\hline
   $Z_{\mu}^{0}\hat{f}^{u\dagger}_LK^{\bar{u}\bar{u}}_L\bar{\sigma}^{\mu}\hat{f}^u_L$ & $V^{u\dagger}_LS^{\bar{u}\bar{u}}_L V^u_L$\\\hline
   $Z_{\mu}^{0}\hat{f}^{u\dagger}_RK^{\bar{u}\bar{u}}_R\bar{\sigma}^{\mu}\hat{f}^u_R$ & $V^{u\dagger}_RD^{\bar{u}\bar{u}}_R V^u_R$\\\hline
   $Z_{\mu}^{0}\hat{f}^{u\dagger}_RK^{uu}_R\bar{\sigma}^{\mu}\hat{f}^u_R$ & $V^{u\dagger}_RS^{uu}_R V^u_R$\\\hline
   $Z_{\mu}^{0}\hat{f}^{d\dagger}_LK^{dd}_L\bar{\sigma}^{\mu}\hat{f}^d_L$ & $V^{d\dagger}_LD^{dd}_L V^d_L$\\\hline
   $Z_{\mu}^{0}\hat{f}^{d\dagger}_RK^{\bar{d}\bar{d}}_R\bar{\sigma}^{\mu}\hat{f}^d_R$ & $V^{d\dagger}_RD^{\bar{d}\bar{d}}_R V^d_R$\\\hline
   $Z_{\mu}^{0}\hat{f}^{d\dagger}_RK^{dd}_R\bar{\sigma}^{\mu}\hat{f}^d_R$ & $V^{d\dagger}_RS^{dd}_R V^d_R$\\\hline
\end{tabular}
\caption{We give the form of all the generalized CKM matrices $K^{ab}_{\alpha}$ and their corresponding triple interaction term coupling the vector bosons $V_{\mu}=W_{\mu},Z_{\mu}, A_{\mu}$ to the quarks in the mass basis (see equation \ref{eq:Lagrangian}). Here, $\hat{f}^{a}_{\alpha}$ are the quark vectors in equation \ref{eq:fs}, and $a,b=u,\bar{u},d,\bar{d}$ and $\alpha=L,R$. The projection matrices $D^{ab}_{\alpha}$ and $S^{ab}_{\alpha}$ are listed in appendix \ref{app:DsAndSs}. The mixing matrices $V^a_{\alpha}$ were calculated numerically and depend on the parameters of the model.}
\label{Tab:CKMmatricesF}
\end{table}
\begin{table}
\renewcommand*\arraystretch{1.0}
\begin{tabular}{|c|c|c|}
\hline
    $\mathbf{V_{\mu}\hat{\tilde{f}}^{a\dagger}\tilde{K}^{ab}_{\alpha}\overleftrightarrow{\partial}^{\mu}\hat{\tilde{f}}^b}$ &
    $\mathbf{V_{\mu}V^{\mu}\hat{\tilde{f}}^{a\dagger}\tilde{K}^{ab}_{\alpha}\hat{\tilde{f}}^b}$ & $\mathbf{\tilde{K}^{ab}_{\alpha}}$\\\hline\hline
    $W_{\mu}^{+}\hat{\tilde{f}}^{u\dagger}\tilde{K}^{ud}_L\overleftrightarrow{\partial}^{\mu}\hat{\tilde{f}}^d$ &
    $W_{\mu}^{+}W^{\mu+}\hat{\tilde{f}}^{u\dagger}\tilde{K}^{ud}_L\hat{\tilde{f}}^d$ & $W^{u\dagger}\tilde{D}^{ud}_L W^d$ \\\hline
    $W_{\mu}^{+}\hat{\tilde{f}}^{d\dagger}\tilde{K}^{\bar{u}\bar{d}\dagger}_R\overleftrightarrow{\partial}^{\mu}\hat{\tilde{f}}^u$ &
    $W_{\mu}^{+}W^{\mu+}\hat{\tilde{f}}^{d\dagger}\tilde{K}^{\bar{u}\bar{d}\dagger}_R\hat{\tilde{f}}^u$ & $W^{u\dagger}\tilde{D}^{\bar{u}\bar{d}}_R W^d$ \\\hline
    $Z_{\mu}^{0}\hat{\tilde{f}}^{u\dagger}\tilde{K}^{uu}_L\overleftrightarrow{\partial}^{\mu}\hat{\tilde{f}}^u$ &
    $Z_{\mu}^{0}Z^{\mu0}\hat{\tilde{f}}^{u\dagger}\tilde{K}^{uu}_L\hat{\tilde{f}}^u$ & $W^{u\dagger}\tilde{D}^{uu}_L W^u$ \\\hline
    $Z_{\mu}^{0}\hat{\tilde{f}}^{u\dagger}\tilde{K}^{\bar{u}\bar{u}}_L\overleftrightarrow{\partial}^{\mu}\hat{\tilde{f}}^u$ &
    $Z_{\mu}^{0}Z^{\mu0}\hat{\tilde{f}}^{u\dagger}\tilde{K}^{\bar{u}\bar{u}}_L\hat{\tilde{f}}^u$ & $W^{u\dagger}\tilde{S}^{\bar{u}\bar{u}}_L W^u$ \\\hline
    $Z_{\mu}^{0}\hat{\tilde{f}}^{u\dagger}\tilde{K}^{\bar{u}\bar{u}}_R\overleftrightarrow{\partial}^{\mu}\hat{\tilde{f}}^u$ &
    $Z_{\mu}^{0}Z^{\mu0}\hat{\tilde{f}}^{u\dagger}\tilde{K}^{\bar{u}\bar{u}}_R\hat{\tilde{f}}^u$ & $W^{u\dagger}\tilde{D}^{\bar{u}\bar{u}}_R W^u$ \\\hline
    $Z_{\mu}^{0}\hat{\tilde{f}}^{u\dagger}\tilde{K}^{uu}_R\overleftrightarrow{\partial}^{\mu}\hat{\tilde{f}}^u$ &
    $Z_{\mu}^{0}Z^{\mu0}\hat{\tilde{f}}^{u\dagger}\tilde{K}^{uu}_R\hat{\tilde{f}}^u$ & $W^{u\dagger}\tilde{S}^{uu}_R W^u$ \\\hline
    $Z_{\mu}^{0}\hat{\tilde{f}}^{d\dagger}\tilde{K}^{dd}_L\overleftrightarrow{\partial}^{\mu}\hat{\tilde{f}}^d$ &
    $Z_{\mu}^{0}Z^{\mu0}\hat{\tilde{f}}^{d\dagger}\tilde{K}^{dd}_L\hat{\tilde{f}}^d$ & $W^{d\dagger}\tilde{D}^{dd}_L W^d$ \\\hline
    $Z_{\mu}^{0}\hat{\tilde{f}}^{d\dagger}\tilde{K}^{\bar{d}\bar{d}}_R\overleftrightarrow{\partial}^{\mu}\hat{\tilde{f}}^d$ &
    $Z_{\mu}^{0}Z^{\mu0}\hat{\tilde{f}}^{d\dagger}\tilde{K}^{\bar{d}\bar{d}}_R\hat{\tilde{f}}^d$ & $W^{d\dagger}\tilde{D}^{\bar{d}\bar{d}}_R W^d$ \\\hline
    $Z_{\mu}^{0}\hat{\tilde{f}}^{d\dagger}\tilde{K}^{dd}_R\overleftrightarrow{\partial}^{\mu}\hat{\tilde{f}}^d$ &
    $Z_{\mu}^{0}Z^{\mu0}\hat{\tilde{f}}^{d\dagger}\tilde{K}^{dd}_R\hat{\tilde{f}}^d$ & $W^{d\dagger}\tilde{S}^{dd}_R W^d$  \\
   \hline
\end{tabular}
\caption{We give the form of all the generalized CKM matrices $\tilde{K}^{ab}_{\alpha}$ and their corresponding triple and quartic interaction term coupling the vector bosons $V_{\mu}=W_{\mu},Z_{\mu}, A_{\mu}$ to the squarks in the mass basis (see equation \ref{eq:LagrangianS}). Here, $\hat{\tilde{f}}^{a}$ are the squark vectors in equation \ref{eq:fs}, and $a,b=u,\bar{u},d,\bar{d}$ and $\alpha=L,R$. The projection matrices $\tilde{D}^{ab}_{\alpha}$ and $\tilde{S}^{ab}_{\alpha}$ are listed in appendix \ref{app:DsAndSs}. The mixing matrices $W^a$ were calculated numerically and depend on the parameters of the model.}
\label{Tab:CKMmatricesS}
\end{table}

As an example, let us write down in matrix form the term in the Lagrangian corresponding to the charged current interaction vertex $W^{+}ff$. In terms of the gauge eigenstate basis vectors $f^{u\dagger}_L$ (a 3-dimensional row vector in generation space) and $f^{d\dagger}_L$ (a 2-dimensional column vector in generation space), the interaction term needs a $3\times2$  projection matrix, which we call $D^{ud}_L$, to couple the L.H fields with $T_3=1/2$ ($u^{c}_3$ and $u^{c}_4$) in $f^{u\dagger}_L$ to the left-handed fields with $T_3=-1/2$ ($d_3$ and $d_4$) in $f^d_L$. This gives a term $\propto W^{+}_{\mu}f_L^{u\dagger}D^{ud}_L\bar{\sigma}^{\mu}f^d_L$. Similarly, in terms of the gauge eigenstate basis vectors $f^{d\dagger}_R$ (a 2-dimensional row vector in generation space) and $f^{u}_R$ (a 3-dimensional column vector in generation space), the interaction term needs a $2\times3$ projection matrix,  $D^{\bar{u}\bar{d}}_R$, to couple the R.H field with $T_3=1/2$ ($\bar{d}_4$) in $f^{d\dagger}_R$ to the right-handed field with $T_3=-1/2$ ($\bar{u}^c_4$) in $f^u_R$. This gives a new term $\propto W^{+}_{\mu}f_R^{d\dagger}D^{\bar{u}\bar{d}\dagger}_R\bar{\sigma}^{\mu}f^u_R$ that is not in the MSSM which couples R.H fields to the $W$ boson. After rotating to the mass eigenstate basis and including the couplings, we get
\begin{equation}
-\mathcal{L}_{W^{+}ff} = \frac{g}{\sqrt{2}}W^{+}_{\mu}\hat{f}^{u\dagger}_{L}K^{ud}_L\bar{\sigma}^{\mu}\hat{f}^d_L + \frac{g}{\sqrt{2}}W^{+}_{\mu}\hat{f}^{d\dagger}_{R}K^   {\bar{u}\bar{d}\dagger}_R\bar{\sigma}^{\mu}\hat{f}^u_R
\end{equation}
from which the coupling matrix   $G^W_{ud}=\frac{g}{\sqrt{2}} K^{ud}_L$ and $G^W_{\bar{u}\bar{d}}= -\frac{g}{\sqrt{2}} K^{\bar{u}\bar{d}}_R$ can be extracted. We give the explicit form of the coupling matrices in Table \ref{Tab:CouplingFerm}, Table \ref{Tab:CouplingScalTrip} and Table \ref{Tab:CouplingScalQuartic}.
\begin{table}
\renewcommand*\arraystretch{1.0}
\center
\begin{tabular}{|c|c|}
  \hline
  \textbf{Coupling Matrix} & \textbf{Explicit Form}\\\hline\hline
 $G^W_{ud}$ & $\frac{g}{\sqrt{2}} K^{ud}_L$\\\hline
 $G^Z_{u_L}$ & $g^Z_{(\frac{1}{2},\frac{2}{3})}K^{uu}_L +g^Z_{(0,\frac{2}{3})}K^{\bar{u}\bar{u}}_L$\\\hline
 $G^Z_{d_L}$ & $g^Z_{(-\frac{1}{2},-\frac{1}{3})}K^{dd}_L$\\\hline
 $G^A_{u_L}$ & $g^A_{\frac{2}{3}}[K^{uu}_L + K^{\bar{u}\bar{u}}_L]$\\\hline
 $G^A_{d_L}$ & $g^A_{-\frac{1}{3}}K^{dd}_L$\\\hline
 $G^W_{\bar{u}\bar{d}}$ & $-\frac{g}{\sqrt{2}} K^{\bar{u}\bar{d}}_R$\\\hline
 $G^Z_{u_R}$ & $g^Z_{(0,-\frac{2}{3})}K^{uu}_R + g^Z_{(-\frac{1}{2},-\frac{2}{3})}K^{\bar{u}\bar{u}}_R$\\\hline
 $G^Z_{d_R}$ & $g^Z_{(0,\frac{1}{3})}K^{dd}_R+g^Z_{(\frac{1}{2},\frac{1}{3})}K^{\bar{d}\bar{d}}_R$ \\\hline
 $G^A_{u_R}$ & $g^A_{\frac{2}{3}}[K^{uu}_R + K^{\bar{u}\bar{u}}_R]$\\\hline
 $G^A_{d_R}$ & $g^A_{-\frac{1}{3}}K^{dd}_R$\\\hline
\end{tabular}
\caption{The coupling matrices at the triple vertex between quarks and gauge bosons. We define $g^Z_{(T^3,Q)}=\frac{g}{\cos\theta_W}(T^3-Q\sin^2\theta_W)$, $g^A_{Q}=Qe$}
\label{Tab:CouplingFerm}
\center
\renewcommand*\arraystretch{1.0}
\begin{tabular}{|c|c|}
  \hline
   \textbf{Coupling Matrix}  & \textbf{Explicit Form}\\\hline\hline
  $\tilde{G}^W_{ud}$ & $\frac{g}{\sqrt{2}}\tilde{K}^{ud}_L$\\\hline
  $\tilde{G}^Z_{u}$ & $g^Z_{(\frac{1}{2},\frac{2}{3})}\tilde{K}^{uu}_L + g^Z_{(0,\frac{2}{3})}\tilde{K}^{\bar{u}\bar{u}}_L +g^Z_{(0,-\frac{2}{3})}\tilde{K}^{uu}_R + g^Z_{(-\frac{1}{2},-\frac{2}{3})}\tilde{K}^{\bar{u}\bar{u}}_R$\\\hline
  $\tilde{G}^A_{u}$ & $g^A_{\frac{2}{3}}\tilde{K}^{uu}_L + g^A_{\frac{2}{3}}\tilde{K}^{\bar{u}\bar{u}}_L + g^A_{\frac{-2}{3}}\tilde{K}^{uu}_R + g^A_{\frac{-2}{3}}\tilde{K}^{\bar{u}\bar{u}}_R$\\\hline
  $\tilde{G}^W_{\bar{u}\bar{d}}$ & $-\frac{g}{\sqrt{2}}\tilde{K}^{\bar{u}\bar{d}}_R$\\\hline
  $\tilde{G}^Z_{d}$ & $g^Z_{(-\frac{1}{2},-\frac{1}{3})}\tilde{K}^{dd}_L+g^Z_{(0,\frac{1}{3})}\tilde{K}^{dd}_R
  +g^Z_{(\frac{1}{2},\frac{1}{3})}\tilde{K}^{\bar{d}\bar{d}}_R$\\\hline
  $\tilde{G}^A_{d}$ & $g^A_{-\frac{1}{3}}\tilde{K}^{dd}_L + g^A_{\frac{1}{3}}\tilde{K}^{dd}_R+ g^A_{\frac{1}{3}}\tilde{K}^{\bar{d}\bar{d}}_R$\\\hline
\end{tabular}
\caption{The coupling matrices at the triple vertex between squarks and gauge bosons.
We define $g^Z_{(T^3,Q)}=\frac{g}{\cos\theta_W}(T^3-Q\sin^2\theta_W)$, $g^A_{Q}=Qe$}
\label{Tab:CouplingScalTrip}
\end{table}

\begin{table}
\renewcommand*\arraystretch{1.0}
\center
\begin{tabular}{|c|c|}
  \hline
  \textbf{Coupling Matrix} & \textbf{Explicit Form}\\\hline\hline
  $\tilde{G}^{WW}_{u}$ & $\frac{g^2}{2}[\tilde{K}^{uu}_L + \tilde{K}^{\bar{u}\bar{u}}_R]$\\\hline
  $\tilde{G}^{WW}_{d}$ & $\frac{g^2}{2}[\tilde{K}^{dd}_L + \tilde{K}^{\bar{d}\bar{d}}_R]$\\\hline
  $\tilde{G}^{ZZ}_{u}$ & $(g^Z_{(\frac{1}{2},\frac{2}{3})})^2\tilde{K}^{uu}_L + (g^Z_{(0,\frac{2}{3})})^2\tilde{K}^{\bar{u}\bar{u}}_L +(g^Z_{(0,-\frac{2}{3})})^2\tilde{K}^{uu}_R + (g^Z_{(-\frac{1}{2},-\frac{2}{3})})^2\tilde{K}^{\bar{u}\bar{u}}_R$\\\hline
  $\tilde{G}^{ZZ}_{d}$ & $(g^Z_{(-\frac{1}{2},-\frac{1}{3})})^2\tilde{K}^{dd}_L+(g^Z_{(0,\frac{1}{3})})^2\tilde{K}^{dd}_R
  +(g^Z_{(\frac{1}{2},\frac{1}{3})})^2\tilde{K}^{\bar{d}\bar{d}}_R$\\\hline
  $\tilde{G}^{ZA}_{u}$ & $2[g^A_{\frac{2}{3}}g^Z_{(\frac{1}{2},\frac{2}{3})}\tilde{K}^{uu}_L + g^A_{-\frac{2}{3}}g^Z_{(0,\frac{2}{3})}\tilde{K}^{\bar{u}\bar{u}}_L +g^A_{\frac{2}{3}}g^Z_{(0,-\frac{2}{3})}\tilde{K}^{uu}_R + g^A_{-\frac{2}{3}}g^Z_{(-\frac{1}{2},-\frac{2}{3})}\tilde{K}^{\bar{u}\bar{u}}_R]$\\\hline
  $\tilde{G}^{ZA}_{d}$ & $2[g^A_{-\frac{1}{3}}g^Z_{(-\frac{1}{2},-\frac{1}{3})}\tilde{K}^{dd}_L+g^A_{-\frac{1}{3}}g^Z_{(0,\frac{1}{3})}\tilde{K}^{dd}_R
  +g^A_{\frac{1}{3}}g^Z_{(\frac{1}{2},\frac{1}{3})}\tilde{K}^{\bar{d}\bar{d}}_R]$\\\hline
  $\tilde{G}^{AA}_{u}$ & $2(g^A_{\frac{2}{3}})^2[\tilde{K}^{uu}_L + \tilde{K}^{\bar{u}\bar{u}}_L +\tilde{K}^{uu}_R +\tilde{K}^{\bar{u}\bar{u}}_R]$\\\hline
  $\tilde{G}^{AA}_{d}$ & $2(g^A_{\frac{1}{3}})^2[\tilde{K}^{dd}_L + \tilde{K}^{dd}_R + \tilde{K}^{\bar{d}\bar{d}}_R]$\\\hline
\end{tabular}
\caption{The coupling matrices at the quartic vertex between squarks and gauge bosons. We define $g^Z_{(T^3,Q)}=\frac{g}{\cos\theta_W}(T^3-Q\sin^2\theta_W)$, $g^A_{Q}=Qe$}
\label{Tab:CouplingScalQuartic}
\end{table}

Proceeding similarly to the above example, the interaction Lagrangian for gauge bosons, quarks and the Higgs in the mass eigenstate basis is:
\begin{align}\label{eq:Lagrangian}
-\mathcal{L}_f &= \space W^{+}_\mu(\hat{f}^{u\dagger}_{L}G^W_{ud}\bar{\sigma}^{\mu}\hat{f}^d_L+\hat{f}^{d\dagger}_{R}G^{W\dagger}_{\bar{u}\bar{d}}\bar{\sigma}^{\mu}\hat{f}^u_R) + W^{-}_\mu(\hat{f}^{d\dagger}_{L}G^{W\dagger}_{ud}\bar{\sigma}^{\mu}\hat{f}^u_L+\hat{f}^{u\dagger}_{R}G^W_{\bar{u}\bar{d}}\bar{\sigma}^{\mu}\hat{f}^d_R)\\
\nonumber&+ Z^{0}_\mu(\hat{f}^{u\dagger}_{L}G^Z_{u_L}\bar{\sigma}^{\mu}\hat{f}^u_L +\hat{f}^{d\dagger}_{L}G^Z_{d_L}\bar{\sigma}^{\mu}\hat{f}^d_L +\hat{f}^{u\dagger}_{R}G^Z_{u_R}\bar{\sigma}^{\mu}\hat{f}^u_R +\hat{f}^{d\dagger}_{R}G^Z_{d_R}\bar{\sigma}^{\mu}\hat{f}^d_R) \\
\nonumber&+ A_\mu(\hat{f}^{u\dagger}_{L}G^A_{u_L}\bar{\sigma}^{\mu}\hat{f}^u_L +\hat{f}^{d\dagger}_{L}G^A_{d_L}\bar{\sigma}^{\mu}\hat{f}^d_L +\hat{f}^{u\dagger}_{R}G^A_{u_R}\bar{\sigma}^{\mu}\hat{f}^u_R +\hat{f}^{d\dagger}_{R}G^A_{d_R}\bar{\sigma}^{\mu}\hat{f}^d_R)\\
\nonumber&+ (h_o \hat{f}_L^{uT} Y^{u\bar{u}} \hat{f}_R^u + h_o \hat{f}_L^{dT} Y^{d\bar{d}} \hat{f}_R^d + \text{h.c})
\end{align}
where $Y^{u\bar{u}}=V^{u\dagger}_L y^{u\bar{u}} V^u_R$ and $Y^{d\bar{d}}=V^{d\dagger}_L y^{d\bar{d}} V^d_R$, with $y^{ab}$ defined as in appendix \ref{app:DsAndSs}, are the matrices coupling the scalar Higgs to the quarks. Similarly, the interaction Lagrangian for gauge bosons and squarks in the mass eigenstate basis is:
\begin{align}\label{eq:LagrangianS}
-\mathcal{L}_{\tilde{f}} &= W^{+}_\mu(\hat{\tilde{f}}^{u\dagger}\tilde{G}^{W}_{ud}\overleftrightarrow{\partial}^{\mu}\hat{\tilde{f}}^d
+ \hat{\tilde{f}}^{d\dagger}\tilde{G}^{W\dagger}_{\bar{u}\bar{d}}\overleftrightarrow{\partial}^{\mu}\hat{\tilde{f}}^u)
+ W^{-}_\mu(\hat{\tilde{f}}^{d\dagger}\tilde{G}^{W\dagger}_{ud}\overleftrightarrow{\partial}^{\mu}\hat{\tilde{f}}^u
+ \hat{\tilde{f}}^{u\dagger}\tilde{G}^{W}_{\bar{u}\bar{d}}\overleftrightarrow{\partial}^{\mu}\hat{\tilde{f}}^d)\\
\nonumber&+ Z^{0}_\mu(\hat{\tilde{f}}^{u\dagger}\tilde{G}^{Z}_{u}\overleftrightarrow{\partial}^{\mu}\hat{\tilde{f}}^u +\hat{\tilde{f}}^{d\dagger}\tilde{G}^{Z}_{d}\overleftrightarrow{\partial}^{\mu}\hat{\tilde{f}}^d )
+ A_\mu(\hat{\tilde{f}}^{u\dagger}\tilde{G}^{A}_{u}\overleftrightarrow{\partial}^{\mu}\hat{\tilde{f}}^u +\hat{\tilde{f}}^{d\dagger}\tilde{G}^{A}_{d}\overleftrightarrow{\partial}^{\mu}\hat{\tilde{f}}^d )\\
\nonumber&+ W^{+}_{\mu} W^{-\mu}(\hat{\tilde{f}}^{u\dagger}\tilde{G}^{WW}_{u}\hat{\tilde{f}}^u +\hat{\tilde{f}}^{d\dagger}\tilde{G}^{WW}_{d}\hat{\tilde{f}}^d)
+Z^{0}_\mu Z^{0\mu}(\hat{\tilde{f}}^{u\dagger}\tilde{G}^{ZZ}_{u}\hat{\tilde{f}}^u + \hat{\tilde{f}}^{d\dagger}\tilde{G}^{ZZ}_{d}\hat{\tilde{f}}^d) \\
\nonumber&+ Z^{0}_\mu A^{\mu}(\hat{\tilde{f}}^{u\dagger}\tilde{G}^{ZA}_{u}\hat{\tilde{f}}^u
+ \hat{\tilde{f}}^{d\dagger}\tilde{G}^{ZA}_{d}\hat{\tilde{f}}^d)
+ A_\mu A^{\mu}(\hat{\tilde{f}}^{u\dagger}\tilde{G}^{AA}_{u}\hat{\tilde{f}}^u
+\hat{\tilde{f}}^{d\dagger}\tilde{G}^{AA}_{d}\hat{\tilde{f}}^d)
\end{align}.

\section{Projection Matrices}\label{app:DsAndSs}
Below, we write down explicitly all of the projection matrices $D^{ab}_{\alpha}$, $S^{ab}_{\alpha}$, $\tilde{D}^{ab}_{\alpha}$ and $\tilde{S}^{ab}_{\alpha}$ used in the construction of the generalized CKM matrices (see appendix \ref{app:Couplings}). It follows that only $D^{ud}_{L}$ and $D^{\bar{u}\bar{d}}_{R}$ (and $\tilde{D}^{ud}_{L}$ and $\tilde{D}^{\bar{u}\bar{d}}_{R}$) are independent, since all of the other matrices can be obtained from their products. For example, $D^{ud}_L (D^{ud}_L)^\dagger = D^{uu}_L$, $(S^{uu}_L)^{\dagger} S^{uu}_L = S^{\bar{u}\bar{u}}_L$, It also follows that $D^{uu}_L + S^{\bar{u}\bar{u}}_L = 1_{3\times3}$. For completeness, we also include the matrices $y^{ab}\subset Y^{ab}$ present in the interaction term coupling the Higgs scalar particle to all third and fourth generation quarks (see \ref{eq:Lagrangian}).
\\
\paragraph{Quark Sector:}
\begin{eqnarray*}
D^{ud}_L&=& \left(
\begin{array}{ccc}
1 & 0 \\
0 & 1 \\
0 & 0 \\
\end{array}
\right).\quad\text{Couples ($T_3=\frac{-1}{2}$) $u^{\dagger}_3,u^{\dagger}_4\in f^{u\dagger}_L$ to ($T_3=\frac{1}{2}$) $d_3,d_4\in f^d_L$.}\\
\end{eqnarray*}
\begin{eqnarray*}
D^{\bar{u}\bar{d}}_{R}&=& \left(
\begin{array}{ccc}
0 & 0 \\
0 & 0 \\
0 & 1 \\
\end{array}
\right).\quad\text{Couples ($T_3=\frac{-1}{2}$) $\bar{d}^{c\dagger}_4\in f^{d\dagger}_R$ to ($T_3=\frac{1}{2}$) $\bar{u}^c_4\in f^u_R$.}\\
\end{eqnarray*}
From the two matrices above, we can construct:
\begin{itemize}
\item $D^{uu}_L = \text{Diag}(1,1,0)$. Couples ($T_3=\frac{-1}{2}$) $u^{\dagger}_3,u^{\dagger}_4\in f^{u\dagger}_L$ to ($T_3=\frac{1}{2}$) $u_3,u_4\in f^u_L$.

\item $S^{\bar{u}\bar{u}}_{L} = \text{Diag}(0,0,1)$. Couples ($T_3=0$) $\bar{u}^c_4\in f^{u\dagger}_L$ to ($T_3=0$) $\bar{u}_4\in f^u_L$.

\item $S^{uu}_R = \text{Diag}(1,1,0)$. Couples ($T_3=0$) $u^{c\dagger}_3, u^{c\dagger}_4\in f^{u\dagger}_R$ to ($T_3=0$) $u^c_3,u^c_4\in f^u_R$.

\item $D^{\bar{u}\bar{u}}_R = \text{Diag}(0,0,1)$. Couples ($T_3=\frac{1}{2}$) $\bar{u}^{c\dagger}_4\in f^{u\dagger}_R$ to ($T_3=\frac{-1}{2}$) $\bar{u}^c_4\in f^u_R$.

\item $D^{dd}_L = \text{Diag}(1,1)$. Couples ($T_3=\frac{1}{2}$) $d^{\dagger}_3,d^{\dagger}_4\in f^{d\dagger}_L$ to ($T_3=\frac{-1}{2}$) $d_3,d_4\in f^d_L$.

\item $S^{dd}_R = \text{Diag}(1,0)$. Couples ($T_3=0$) $d^{c\dagger}_3\in f^{d\dagger}_R$ to ($T_3=0$) $d^c_3\in f^d_R$.

\item $D^{\bar{d}\bar{d}}_R = \text{Diag}(0,1)$. Couples ($T_3=\frac{-1}{2}$) $\bar{d}^{c\dagger}_4\in f^{u\dagger}_R$ to ($T_3=\frac{1}{2}$) $\bar{d}^c_4\in f^d_R$.
\end{itemize}

\paragraph{Squark Sector:}
\begin{eqnarray*}
\tilde{D}^{ud}_L&=& \left(
\begin{array}{cccccc}
1 & 0 & 0 & 0  \\
0 & 1 & 0 & 0  \\
0 & 0 & 0 & 0  \\
0 & 0 & 0 & 0  \\
0 & 0 & 0 & 0  \\
0 & 0 & 0 & 0  \\
\end{array}
\right).\qquad \text{Couples ($T_3=\frac{-1}{2}$) $\tilde{u}^{*}_3, \tilde{u}^{*}_4\in \tilde{f}^{u\dagger}$ to ($T_3=\frac{1}{2}$) $\tilde{d}_3, \tilde{d}_4\in \tilde{f}^d$.}\\
\end{eqnarray*}
\begin{eqnarray*}
\tilde{D}^{\bar{u}\bar{d}}_R&=& \left(
\begin{array}{cccccc}
0 & 0 & 0 & 0  \\
0 & 0 & 0 & 0  \\
0 & 0 & 0 & 0  \\
0 & 0 & 0 & 0  \\
0 & 0 & 0 & 0  \\
0 & 0 & 0 & 1  \\
\end{array}
\right).\qquad \text{Couples ($T_3=\frac{-1}{2}$) $\tilde{\bar{d}}^{c*}_4\in \tilde{f}^{d\dagger}$ to ($T_3=\frac{1}{2}$) $\tilde{\bar{u}}^c_4\in \tilde{f}^u$.}\\
\end{eqnarray*}
From the two matrices above, we can construct:
\begin{itemize}
\item $\tilde{D}^{uu}_L = \text{Diag}(1,1,0,0,0,0)$. Couples ($T_3=\frac{-1}{2}$) $\tilde{u}^{*}_3,\tilde{u}^{*}_4\in \tilde{f}^{u\dagger}$ to ($T_3=\frac{1}{2}$) $u_3,u_4\in \tilde{f}^u$.

\item $\tilde{S}^{\bar{u}\bar{u}}_{L} = \text{Diag}(0,0,1,0,0,0)$. Couples ($T_3=0$) $\tilde{\bar{u}}^c_4\in \tilde{f}^{u\dagger}$ to ($T_3=0$) $\bar{u}_4\in \tilde{f}^u$.

\item $\tilde{S}^{uu}_{R} = \text{Diag}(0,0,0,1,1,0)$. Couples ($T_3=0$) $\tilde{u}^{c*}_3, \tilde{u}^{c*}_4\in \tilde{f}^{u\dagger}$ to ($T_3=0$) $\tilde{u}^c_3,\tilde{u}^c_4\in \tilde{f}^u$.

\item $\tilde{D}^{\bar{u}\bar{u}}_{R} = \text{Diag}(0,0,0,0,0,1)$. Couples ($T_3=\frac{1}{2}$) $\tilde{\bar{u}}^{c*}_4\in \tilde{f}^{u\dagger}$ to ($T_3=\frac{-1}{2}$) $\tilde{\bar{u}}^c_4\in \tilde{f}^u$.

\item $\tilde{D}^{dd}_{L} = \text{Diag}(1,1,0,0)$.Couples ($T_3=\frac{1}{2}$) $\tilde{d}^{*}_3,\tilde{d}^{*}_4\in \tilde{f}^{d\dagger}$ to ($T_3=\frac{-1}{2}$) $\tilde{d}_3, \tilde{d}_4\in \tilde{f}^d$.

\item $\tilde{S}^{dd}_{R} = \text{Diag}(0,0,1,0)$. Couples ($T_3=0$) $\tilde{d}^{c*}_3\in \tilde{f}^{d\dagger}$ to ($T_3=0$) $\tilde{d}^c_3\in \tilde{f}^d$.

\item $\tilde{D}^{\bar{d}\bar{d}}_{R} = \text{Diag}(0,0,0,1)$. Couples ($T_3=\frac{-1}{2}$) $\tilde{\bar{d}}^{c*}_4\in \tilde{f}^{u\dagger}$ to ($T_3=\frac{1}{2}$) $\tilde{\bar{d}}^c_4\in \tilde{f}^d$.
\end{itemize}

\paragraph{Higgs Sector:}
\begin{equation*}
y^{u\bar{u}} = \left(
\begin{array}{ccc}
y_{33} & y_{34} & 0  \\
y_{43} & y_{44} & 0  \\
0 & 0 & 0  \\
\end{array}
\right)
\quad\text{and}\quad
y^{d\bar{d}} = \left(
\begin{array}{cc}
y_\text{bot} & 0  \\
0 & 0  \\
\end{array}
\right).\\
\end{equation*}

\section{Beta Functions}\label{app:BetaFunctions}
\paragraph*{Gauge Couplings:}\label{app:BetaFunctionsGauge}
The beta function for the gauge couplings are:
\begin{equation*}
16\pi^2 \frac{dg_i}{dt}=-b_i g_i^3.
\end{equation*}
Here, $t=\ln Q$ where $Q$ is the renormalization scale. The beta function coefficients for an arbitrary number of SU(5) multiplets $n_5$ and $n_{10}$ are given by:
\begin{align*}
b1 &= \frac{3}{5}(11) + n_{10} b_{10} + n_5  b_5 \\
b2 &= 1 + n_{10} b_{10}  + n_5 b_5 \\
b3 &=  -3 +  n_{10}  b_{10} + n_5 b_5
\end{align*}
with $b_{10} = 3$, $b_5 = 1$ denoting group theoretic coefficients.
\\
\paragraph*{Top Yukawa Coupling:}\label{app:BetaFunctionsYuk}
\indent Using the general results in \cite{Martin:1993zk}, we obtain the following top Yukawa two-loop beta function:
\begin{align*}
\beta_{Y_u}(t) &= \frac{1}{16 \pi^2} \Bigg( (3 \text{Tr}[Y_u(t).Y^{\dagger}_u(t)] Y_u(t)
+ 3 Y_u(t)Y^{\dagger}_u(t)Y_u(t) \\
\nonumber &+ Y_u(t) Y^{\dagger}_d(t)Y_d(t))- (\frac{16}{3} g_3(t)^2 + 3 g_2(t)^2 + \frac{13}{15} g_1(t)^2) Y_u(t) \Bigg).
\end{align*}
Here, $Y_u$ is the up-type Yukawa coupling matrix containing $y_{33}$, $y_{34}$, $y_{43}$ and $y_{44}$.

\section{Calculation of Oblique Parameters}\label{app:SandT}
\paragraph*{Fermion Contribution:}
\indent In \cite{Lavoura:1992np}, the authors derived a general formula for computing the values of $S$ and $T$ for any model with vector-like quarks, where the number of up and down quarks are arbitrary and not necessarily equal. Adapting these general results to our model, we get:
\begin{align*}
T_{f} &= \frac{3}{16 \pi \sin^2\theta_W\cos^2\theta_W} \Bigg( \sum^3_{\alpha=1} \sum^2_{i=2}\left( [(K^{ud}_L)_{\alpha i}^2+(K^{\bar{u}\bar{d}}_R)_{\alpha i}^2]\theta_{+}(y_\alpha, y_i) + 2[(K^{ud}_L)_{\alpha i}(K^{\bar{u}\bar{d}}_R)_{\alpha i}]\theta_{-}(y_\alpha, y_i) \right) \\
&- \sum_{\beta<\alpha}\left([(K^{uu}_L)_{\alpha\beta}^2+(K^{\bar{u}\bar{u}}_R)_{\alpha\beta}^2]\theta_{+}(y_\alpha, y_\beta) + 2[(K^{uu}_L)_{\alpha\beta}(K^{\bar{u}\bar{u}}_R)_{\alpha\beta}]\theta_{-}(y_\alpha, y_\beta) \right) \\
&- \sum_{j<i}\left([(K^{dd}_L)_{ij}^2+(K^{\bar{d}\bar{d}}_R)_{ij}^2]\theta_{+}(y_i, y_j) + 2[(K^{dd}_L)_{ij}(K^{\bar{d}\bar{d}}_R)_{ij}]\theta_{-}(y_i, y_j) \right) \Bigg) , \\
S_{f} &= \frac{3}{2 \pi} \Bigg( \sum^3_{\alpha=1} \sum^2_{i=2}\left([K^{ud}_L)_{\alpha i}^2+(K^{\bar{u}\bar{d}}_R)_{\alpha i}^2]\psi_{+}(y_\alpha, y_i) + 2[(K^{ud}_L)_{\alpha i}(K^{\bar{u}\bar{d}}_R)_{\alpha i}]\psi_{-}(y_\alpha, y_i)\right) \\
&- \sum_{\beta<\alpha}\left([(K^{uu}_L)_{\alpha\beta}^2+(K^{\bar{u}\bar{u}}_R)_{\alpha\beta}^2]\chi_{+}(y_\alpha, y_\beta) + 2[(K^{uu}_L)_{\alpha\beta}(K^{\bar{u}\bar{u}}_R)_{\alpha\beta}]\chi_{-}(y_\alpha, y_\beta)\right) \\
&- \sum_{j<i}\left([(K^{dd}_L)_{ij}^2+(K^{\bar{d}\bar{d}}_R)_{ij}^2]\chi_{+}(y_i, y_j) + 2[(K^{dd}_L)_{ij}(K^{\bar{d}\bar{d}}_R)_{ij}]\chi_{-}(y_i, y_j)\right) \Bigg)
\end{align*}
where the $K$'s are the generalized CKM matrices for fermions, derived in appendix \ref{app:Couplings}. The Greek indices sum over the up-type quark generations (i.e from 1 to 3) and the Latin indices sum over the number of down-type quark generations (i.e from 1 to 2). The functions $\theta_{\pm}(y_1, y_2)$, $\psi_{\pm}(y_1, y_2)$ and $\chi_{\pm}(y_1, y_2)$ are defined in appendix \ref{app:functions}, and $y_i\equiv m^2_i/m^2_Z$.
\\
\paragraph*{Scalar Contribution:}
\indent The scalar partners also contribute to the oblique corrections.
For this calculation, we use the notation and conventions of \cite{EidelmanPDG}, where the oblique parameters $S$ and $T$ are defined as
\begin{align*}
S_{s} &= \frac{4\sin^2\theta_W \cos^2\theta_W}{\alpha m^2_Z}\left(\Pi_{ZZ}(m^2_Z)-\Pi_{ZZ}(0)-\frac{\cos^2\theta_W}{\cos\theta_W\sin\theta_W}\Pi_{Z\gamma}(m^2_Z)-\Pi_{\gamma\gamma}(m^2_Z)\right) \\
T_{s} &= \frac{1}{\alpha}\left(\frac{\Pi_{WW}(0)}{m^2_W} - \frac{\Pi_{ZZ}(0)}{m^2_Z}\right)
\label{Eqn:SandTdef}
\end{align*}
where the $\Pi$'s are the electroweak vector boson self-energies. The contributions to the self-energies of the vector bosons from the additional scalars $\tilde{t}'_{1,2,3,4}$ and $\tilde{b}'_{1,2}$ are \cite{Martin:2005SandT}:

\begin{align*}
\Delta\Pi_{\gamma\gamma} &= \frac{3}{16\pi^2}g^2\sin^2\theta_W\left( \left(\frac{2}{3}\right)^2\sum_{i=3}^{6} F(\tilde{t}'_i,\tilde{t}'_i) + \left(\frac{1}{3}\right)^2\sum_{i=3}^{4} F(\tilde{b}'_i,\tilde{b}'_i)\right) \\
\Delta\Pi_{Z\gamma} &= \frac{3}{16\pi^2}g\sin\theta_W\left(\frac{2}{3}\sum_{i=3}^{6} (\tilde{G}^{Z}_u)_{ii} F(\tilde{t}'_i,\tilde{t}'_i) + \frac{1}{3}\sum_{i=3}^{4} (\tilde{G}^{Z}_d)_{ii} F(\tilde{b}'_i,\tilde{b}'_i)\right) \\
\Delta\Pi_{ZZ} &= \frac{3}{16\pi^2}\left(\sum_{i,j=3}^{6} |(\tilde{G}^{Z}_u)_{ij}|^2 F(\tilde{t}'_i,\tilde{t}'_j) + \sum_{i,j=3}^{4} |(\tilde{G}^{Z}_d)_{ij}|^2 F(\tilde{b}'_i,\tilde{b}'_j)\right) \\
\Delta\Pi_{WW} &= \frac{3}{16\pi^2}\sum_{i=3}^{6}\sum_{j=3}^{4} |(\tilde{G}^{W}_{ud})_{ij}|^2 F(\tilde{b}'_i,\tilde{t}'_j)
\end{align*}
where the $\tilde{G}$'s are the coupling matrices for scalars derived in appendix \ref{app:Couplings} and the function $F(x,y)$ is given in appendix \ref{app:functions}.

\section{Useful Functions}\label{app:functions}
The expressions for $\theta_{\pm}(y_i,y_j)$, $\psi_{\pm}(y_i,y_j)$ and $\chi_{\pm}(y_i,y_j)$, used in appendix \ref{app:SandT} are \cite{Lavoura:CKM}:
\begin{align*}
\theta_{+}(y_i,y_j)&=y_i+y_j-\frac{2 y_i y_j}{y_i-y_j}\ln\frac{y_i}{y_j}\\
\theta_{-}(y_i,y_j)&=2\sqrt{y_iy_j} \left( \frac{y_i + y_j}{y_i-y_j}\ln\frac{y_i}{y_j}-2 \right)\\
\psi_{+}(y_i,y_j)&=\frac{1}{3}-\frac{1}{9}\ln\frac{y_i}{y_j}\\
\psi_{-}(y_i,y_j)&=-\frac{y_i+y_j}{6\sqrt{y_i y_j}}\\
\chi_{+}(y_i,y_j)&=\frac{5(y_i^2+y_j^2)-22y_iy_j}{9(y_i+y_j)^2} + \frac{3y_1y_2(y_i+y_j)-y_i^3-y_j^3}{3(y_i-y_j)^3}\ln\frac{y_i}{y_j}\\
\chi_{-}(y_i,y_j)&=-\sqrt{y_iy_j} \left( \frac{y_i+y_j}{6y_iy_j}-\frac{y_i+y_j}{(y_i-y_j)^2}+\frac{2y_iy_j}{(y_i+y_j)^3}\ln\frac{y_i}{y_j} \right).
\end{align*}
Here, $y_i=m^2_i/m^2_Z$, $y_i=m^2_i/m^2_Z$ and the limit $\epsilon \rightarrow 0$ of Dimensional Regularization is assumed. The expression for $F(x,y)$ in the self-energy functions in appendix \ref{app:SandT} is \cite{Martin:2005SandT}:
\begin{align*}
F(x,y)&=H(x,y)+(x+y-p^2)B(x,y)\\
H(x,y) &= \Bigg( 2p^2 -x -y - (x-y)^2/p^2 \Bigg)B(x,y)/3 \\
&+ \Bigg( x\bar{\ln}x+m^2_y \bar{\ln}y-p^2/3 + (x\bar{\ln}x- x - y\bar{\ln}y+y)(y-x)/(2p^2) \Bigg)2/3 \\
B(x,y)&= -\int^1_0dt \bar{\ln}\Bigg(tx + (1-t)y-t(1-t)p^2-i\epsilon \Bigg),
\end{align*}
where now $x=m^2_x$, $y=m^2_y$ and $\bar{\ln}X=\ln(X /m^2_Z)$.
\section*{Acknowledgments}
We particularly thank D.~E.~Kaplan, as well as S.~Rajendran,  both of whom made contributions to this work. We also want to thank C.~Brust and M.~Walters for useful comments.

\end{document}